\documentclass[aps,pra,reprint,amsmath,amssymb,nofootinbib,floatfix,superscriptaddress]{revtex4-2}
\usepackage[table]{xcolor}  % Colours  
\definecolor{PRDblue}{HTML}{2E3092}     % Links
\definecolor{active}{HTML}{FDEFD6}      % Table active
\definecolor{reactive}{HTML}{ECF4FF}    % Table reactive
\usepackage{multirow}       % multirow cells
\usepackage{hhline}         % better horizontal line
\usepackage{fixes}          % Fixes many revtex errors
\usepackage[colorlinks=true, allcolors=PRDblue, pdfborder={0 0 0}]{hyperref}
\usepackage{cleveref}       % Clear referencing
\usepackage{orcidlink}
\usepackage{mathtools}      % Math tools for spacing of maths
\usepackage{isomath}        % Provides fonts conforming with ISO norms
\usepackage{esint}          % Extends the types of integral symbols available
\usepackage{siunitx}        % SI units
\usepackage{physics}        % Essential physics package
\providecommand{\ii}{\text{i}}  % Imaginary unit upright (ISO norm)
\providecommand{\ee}{\text{e}}  % Euler's number upright (ISO norm)
\providecommand{\vc}{\vb*}      % Bold italic vectors    (ISO norm)
\providecommand{\uv}{\vu*}      % Unit vector with hat   (ISO norm)
\newcommand{\tens}[1]{\mathsfbfit{{#1}}}
\renewcommand{\Re}{\real}       % Fraktur R (TeX norm)
\renewcommand{\Im}{\imaginary}  % Fraktur I (TeX norm)
\providecommand{\citer}[1]{ref.~\cite{{#1}}}    % ref. [1]
\providecommand{\citers}[1]{refs.~\cite{{#1}}}  % refs. [1,2]
\begin{document}

\title{Dipole light-matter interactions in the bispinor formalism}
\author{Sebastian \surname{Golat}\,\orcidlink{0000-0003-3947-7634}}  
% \email{sebastian.1.golat@kcl.ac.uk}
\affiliation{
Department of Physics and London Centre for Nanotechnology, 
King's College London, Strand, London WC2R 2LS, UK}

\author{Alex J. \surname{Vernon}\,\orcidlink{0000-0002-3741-4202}}    
\affiliation{Donostia International Physics Center (DIPC), Donostia-San Sebasti\'an 20018, Spain}
% \email{alexander.vernon@kcl.ac.uk}

\author{Francisco J. \surname{Rodr\'iguez-Fortu\~no}\,\orcidlink{0000-0002-4555-1186}}
\email{francisco.rodriguez\_fortuno@kcl.ac.uk}

\affiliation{
Department of Physics and London Centre for Nanotechnology, 
King's College London, Strand, London WC2R 2LS, UK}

\date{\today}

\begin{abstract}
The conventional formulation of power absorption, optical forces, and torques on dipolar particles involve lenghty and cumbersome expressions that obscure their shared physical origin. We apply a  bispinor formalism that unifies these disparate phenomena in a very general case including chiral and nonreciprocal particles. This reveals that force, torque, absorbed power, and absorbed helicity rate can all be concisely expressed in terms of broken symmetries, and leads to the fundamental inequalities that dipolar particles' cross-sections must satisfy. This framework uncovers profound connections normally hidden behind complex algebra—for instance, pressure forces depend exclusively on the difference in linear momenta of different light components and the corresponding breaking of symmetry by a particle, and optical recoil forces depend exclusively on helicity cross sections—providing clarity, conciseness, and a powerful predictive tool for arbitrary dipole interactions.
\end{abstract}

\maketitle
\section{Introduction}
The notion that light waves exist by mutual propagation of an electric and a magnetic field and that these fields must be equally important is fundamental to the form of Maxwell's equations in free space.
This `symmetry of importance' of the two fields is only broken in the presence of matter: typically the dynamics of the electric field is more strongly engaged by atoms, molecules and particles, and as a result it is commonplace in optics to measure only electric quantities like polarisation, energy, momentum, and angular momentum densities. But without considering the combined behaviour of electric and magnetic fields, over space and in their interaction with matter, the more exotic mechanical capabilities of light and their physical significance become obscured. The electromagnetic bispinor, which unites the electric and magnetic fields as vector-valued components of a $\mathbb{C}^2$ vector, offers a compact and dual representation of electromagnetic fields and their dynamical quantities, such as the flow of energy and linear and angular momentum densities, intuitively defined  with fundamental operators  \cite{Alpeggiani2018,Bliokh2014a,Berry2009,Silveirinha2015,Gangaraj2017,Bliokh2014}.

Applying the bispinor formalism to dipolar moments brings intuition and simplicity to dipolar interactions \cite{Bliokh2014}. Perhaps the most powerful and underutilised property of the bispinor is that its $\mathbb{C}^2$ basis can be changed, providing a single unifying framework for all electromagnetic quadratic quantities \cite{golat2024electromagnetic}, surfacing novel electromagnetic wave singularities and topologies \cite{Vernon2025Aug,Vernon2025Jul}, enabling the discovery of a new geometric phase of nonparaxial light \cite{Vernon2026Apr} and leading to vastly simplified optical-force expressions \cite{Golat2023}.

%Perhaps the most powerful and underutilised property of the bispinor is that its $\mathbb{C}^2$ basis can be changed. Not only can a change of bispinor basis greatly simplify some expressions for dipole interactions \cite{Bliokh2014,Golat2023}, but it can unify electromagnetic quantities under a single framework. In a recent work \cite{golat2024electromagnetic}, we leveraged the similarity of the bispinor to Jones vectors of plane wave polarisation to systematically define all quadratic quantities that could appear in dipole interactions in a form analogous to Stokes parameters.

%-----

% Equations like dipole force and dipole absorption, which are usually relatively complicated in the general case of chiral and/or nonreciprocal particles, become elegant and concise in this formalism \cite{Bliokh2014,golat2024electromagnetic}. In this paper, we aim to further leverage this framework to continue this simplifying trend, which started in \cite{Bliokh2014} with the use of dipolar bispinors and operator notation, by further using the bispinor formalism in different bases \cite{golat2024electromagnetic} to express absorption and scattering terms of absorbed power, force, and torque on dipoles, in their full form with no omitted terms, prove fundamental inequalities that  dipolar particles must obey in their cross sections and polarisabilities, both in general or for passive particles, and also present complete and concise equations for a previously omitted `helicity absorption rate' \cite{nieto2015optical} that could prove to be very relevant for experimentalists.

%-------
In this paper, we continue and complete this line of work. We extend the bispinor operator formalism \cite{Bliokh2014} to include the recoil terms that were previously omitted, obtaining compact and fully general expressions for the absorbed power, optical force, and torque on linear dipolar particles in their most general case —including chiral and nonreciprocal cases— expressed concisely within the unifying framework of ref.~\cite{golat2024electromagnetic}. This comprehensive formulation lets us derive general fundamental inequalities that any dipolar particle's cross sections must satisfy, with stricter bounds for passive particles. We also present complete and concise expressions for the overlooked `helicity absorption rate' \cite{nieto2015optical}, a quantity that could prove very relevant for experimentalists, and show that its cross sections are identical to those governing recoil forces—an unexpected connection.

%-------
We start our description with symmetry arguments. Maxwell's equations describing light are highly symmetrical. Any translation, both in time and in space, and rotation leave them invariant; on top of that, in the absence of free charges, applying the so-called duality rotation (rotating electric and magnetic fields into each other) is a symmetry \cite{cameron2012electric, Bliokh2013}. According to Noether's theorem, whenever there is a continuous symmetry, there is a conserved charge associated with it \cite{Noether_1971,Nesterenko1991Sep,Liberal2024Dec}. The translational symmetry leads to the conservation of energy and momentum (for time and space, respectively) and rotational symmetry is associated with the conservation of total angular momentum. Finally, the duality rotation leads to the conservation of helicity (projection of spin onto the direction of motion), which in turn means that with no free charges present, the orbital and spin angular momentum of light have to be separately conserved quantities \cite{Barnett2012,Jajin2024Dec}.
In the absence of matter, the energy, momentum, helicity, and spin of light are conserved quantities. However, when light interacts with a particle, these properties can be transferred to their mechanical analogues. For example, energy conservation, commonly called the optical theorem, states that the extinction power is the sum of the absorbed and scattered power $\langle\mathcal{P}_\text{ext}\rangle=\langle\mathcal{P}_\text{abs}\rangle+\langle\mathcal{P}_\text{sca}\rangle$.
Expressed in SI units, the power absorbed $\langle\mathcal{P}_\text{abs}\rangle=\langle\mathcal{P}_\text{ext}\rangle-\langle\mathcal{P}_\text{sca}\rangle$ by an electric-magnetic dipole is as follows \cite{Toftul2026Apr,NietoVesperinas2010}: 
\begin{equation*}
    \langle\mathcal{P}_\text{abs}\rangle=\underbrace{\tfrac{\omega}{2}\Im\qty(\vc{E}^\ast\!\!\vdot\vc{p}+\mu\vc{H}^\ast\!\!\vdot\vc{m})}_\text{extinction}-\underbrace{\tfrac{\omega k^3}{12\pi}(\tfrac{1}{\varepsilon}\abs{\vc{p}}^2+\mu\abs{\vc{m}}^2)}_\text{scattering},
\end{equation*}
where $\omega$ is the vacuum angular frequency, $\varepsilon$, $\mu$ are the permittivity and permeability of the host medium, which we assume to be real valued throughout the paper, $k=\omega\sqrt{\varepsilon\mu}$ is the wavenumber, ($\vc{p}$, $\vc{m}$) are the induced electric and magnetic dipoles that exist at the same point in space and
($\vc{E}$, $\vc{H}$) are the incident electric and magnetic fields evaluated at that point, all considered as time-harmonic phasors.\footnote{We assume the standard SI units: $[\vc{p}]=\si{\coulomb\meter}$, $[\vc{m}]=\si{\ampere\meter\squared}$, $[\vc{E}]=\si{\volt\per\meter}$, $[\vc{H}]=\si{\ampere\per\meter}$.}
Somewhat recently, it has been shown that a similar optical theorem can be obtained for helicity \cite{nieto2015optical}. This is a quantity that has been overlooked so far. %, but is fundamental and can be very relevant to experiments
Helicity is a conserved quantity that can be absorbed by a dipolar particle at a given rate. Because each handedness of light possesses distinct energy density and flux, this rate corresponds to the difference in power between the right- and left-handed field components (see \cref{app:handedness}).
% constituting a helicity analogue of the Purcell factor of energy extinction in light-matter interactions. 
The rate of helicity absorption $\langle\dot{\mathcal{H}}_\text{abs}\rangle$ (the overdot emphasises that it is a rate of change of helicity) is analogous to power (which is the rate of change of energy) and in SI units it is defined as \cite{nieto2015optical, NietoVesperinas2021}:\footnote{Note that our $\langle\dot{\mathcal{H}}\rangle$ differs from \cite{nieto2015optical, NietoVesperinas2021} by a factor of $4\pi c$ accounting for conversion to SI units of energy $[\dot{\mathcal{H}}]=\si{(\joule\cdot\second)\per\second}=\si{\joule}$.}
\begin{equation*}
    \langle\dot{\mathcal{H}}_\text{abs}\rangle=\underbrace{\tfrac{1}{2}\vphantom{\tfrac{k^3}{6\pi}}\Re(\tfrac{1}{c}\vc{E}^\ast\!\!\vdot\vc{m}-\eta\vc{H}^\ast\!\!\vdot\vc{p})}_\text{extinction}-\underbrace{\tfrac{k^3}{6\pi}\Im(\eta\vc{p}\vdot\vc{m}^\ast)}_\text{scattering},\!\!\!
\end{equation*}
where $c=1/\sqrt{\varepsilon\mu}$ is the phase velocity in the medium. The rate of helicity extinction is again a sum of absorption and scattering terms $\langle{\dot{\mathcal{H}}_\text{ext}\rangle=\langle\dot{\mathcal{H}}_\text{abs}\rangle+\langle\dot{\mathcal{H}}_\text{sca}}\rangle$. Notice that in both cases the extinction is linear in dipoles while the scattering term is quadratic. The linear electromagnetic momentum will be transferred to a dipole via a force which can be expressed as \cite{NietoVesperinas2010,Chaumet2009, Albaladejo2009, Gao2017, Toftul2026Apr}\footnote{The product $\vc{A}\vdot(\grad)\vc{B} \equiv \sum_i ({A}_i \grad {B}_i)$.}
\begin{equation*}
    \langle{\vc{F}}\rangle=\underbrace{\tfrac{1}{2}\Re\big[\vc{p}^\ast\!\!\vdot(\grad)\vc{E}+\mu\vc{m}^\ast\!\!\vdot(\grad)\vc{H}\big]}_\text{interaction}\!\underbrace{-\tfrac{k^4}{12\pi}\Re(\eta\vc{p}\cp\vc{m}^\ast)}_\text{recoil}\,,
\end{equation*}
where $\eta=\sqrt{\mu/\varepsilon}$ is the impedance of the medium.
Notice that there is a similar split into interaction terms linear in dipoles and the quadratic recoil term. {An optical force can indeed be seen as the rate at which the electromagnetic linear momentum is converted into mechanical linear momentum, that is, the rate of absorption of the linear momentum of light, equal to the rate of extinction (interaction term) minus the rate of scattering (recoil term) of the linear momentum.}
Finally, the angular momentum of the electromagnetic field will produce a torque in the dipole \cite{wei2022optical,Toftul2026Apr,Chaumet2009, Canaguier-Durand2013PRA, Bliokh2014a, Bliokh2015, Nieto-Vesperinas2015OL}
\begin{equation*}
    \langle{\vc{\varGamma}}\rangle=\underbrace{\tfrac{1}{2}\Re(\vc{p}^\ast\!\!\cp\!\vc{E}\!+\!\mu\vc{m}^\ast\!\!\cp\!\vc{H})}_\text{interaction}\!
    \underbrace{-\tfrac{ k^3}{12\pi}\Im\big(\tfrac{1}{\varepsilon}\vc{p}^\ast\!\!\cp\!\vc{p}\!+\!\mu\vc{m}^\ast\!\!\cp\!\vc{m}\big)}_\text{recoil}\!.\!\!
\end{equation*}
These equations, which are all derived from very fundamental conservation arguments, are generally applicable as long as the particle is scattering as an electric and magnetic dipole, with no higher order multipole terms, which is always true for small particles in the Rayleigh regime with sizes $R\ll\lambda$, and can be true for larger particles at spectral ranges where the dipolar terms dominate and higher order multipoles can be neglected. The equations work regardless of the nature of the dipoles, including anisotropic, chiral, nonreciprocal, and non-linear responses.

A very common simplification is to assume linearity. If the response of the dipole moments is linear to externally applied fields, one can write them in terms of polarisability tensors ($\tens\alpha_\text{e},\tens\alpha_\text{m},\tens\alpha_\text{c},\tens\alpha_\text{t}$) as \cite{Golat2023,Bliokh2014,Toftul2026Apr,golat2024electromagnetic}
\begin{equation}
    \label{eq:generaltensorpolarisability} 
    \begin{pmatrix}
        \vc{p}/\sqrt{\varepsilon}\\\sqrt{\mu}\vc{m}
    \end{pmatrix}
    =
    \begin{pmatrix}
        \tens\alpha_\text{e}&\tens\alpha_\text{t}+\ii\tens\alpha_\text{c}\\\tens\alpha^\intercal_\text{t}-\ii\tens\alpha^\intercal_{\text{c}\vphantom{\text{t}}}&\tens\alpha_\text{m}
    \end{pmatrix}
    \!
    \begin{pmatrix}
        \sqrt{\varepsilon}\vc{E}\\\sqrt{\mu}\vc{H}
    \end{pmatrix},
\end{equation}
where components of these electric-magnetic vectors are chosen so that they have identical units, e.g., $[\sqrt{\varepsilon}\vc{E}]=[\sqrt{\mu}\vc{H}]=\sqrt{\si{\joule\per\meter\cubed}}$, while all polarisabilities have dimensions of volume, $[\alpha_\text{e}]=[\alpha_\text{m}]=[\alpha_\text{c}]=[\alpha_\text{t}]=\si{\meter\cubed}$. The off-diagonal components are customarily split as a sum of the symmetric component $\tens\alpha_\text{t}$, associated with nonreciprocity and time-symmetry breaking, and the antisymmetric component $\ii\tens\alpha_{\text{c}}$, associated with chirality and parity-symmetry breaking. 
In this work, we will make a further assumption of a bi-isotropic particle response, or alternatively for randomly oriented anisotropic particles, whose polarisability tensor is then averaged over all orientations, thereby reducing the effective polarisability tensors ($\tens\alpha_\text{e},\tens\alpha_\text{m},\tens\alpha_\text{c},\tens\alpha_\text{t}$) to (pseudo)scalar quantities ($\alpha_\text{e},\alpha_\text{m},\alpha_\text{c},\alpha_\text{t}$). Substituting these general polarisabilities into the previous expressions can lead to rather complex expressions in a general anisotropic case \cite{Mun2020,Wang2014,Chen2016,Li2019}. We will show that, combined with the bispinor formalism \cite{Bliokh2014} and the different bases in electric-magnetic space \cite{golat2024electromagnetic}, we can obtain very intuitive expressions for the rate of absorption of energy, helicity, linear momentum, and angular momentum, by a bi-isotropic (or orientation-averaged) linear electromagnetic dipole. These simplified expressions clearly illustrate the symmetry-breaking properties of the dipolar particle and reflect relationships between various terms of power, helicity rate, force, and torque. % We also suggest how these relationships can be utilised to predict orientation-averaged forces and torques for arbitrary dipolar particles just from measuring the extinction power and helicity scattering rate.

% \paragraph*{Bispinor formalism}\!\!\!---\,\,\,%
\section{Bispinor formalism}
The bispinor notation is a well-known mathematical formalism to treat electromagnetic fields \cite{Alpeggiani2018,Bliokh2014,Bliokh2014a}, and the way in which it is specifically used in this work, expressing the electromagnetic bispinor in bases different from the usual electric and magnetic basis, was introduced in detail in \citer{golat2024electromagnetic}. Here is a brief overview of the basic idea: for a time harmonic field, we can define an electromagnetic bispinor. The most usual representation of the electromagnetic bispinor is to denote it as a 6-dimensional column vector $\vc{\psi}(\vc{r})=\tfrac{1}{2}(\sqrt{\varepsilon} \vc{E}(\vc{r}),\sqrt{\mu} \vc{H}(\vc{r}))$, but, like any other vector, we can also express it as an expansion or linear combination of unit vectors:
\begin{equation}\label{eq:bispinor}
    \vc{\psi}(\vc{r})=\tfrac{1}{2} (\sqrt{\varepsilon} \vc{E}(\vc{r}) \otimes\uv{e}_\text{e} +  \sqrt{\mu} \vc{H}(\vc{r})\otimes\uv{e}_\text{m})\,,
\end{equation}
% \begin{equation}\label{eq:bispinor}
%     \vc{\psi}(\vc{r})=\tfrac{\sqrt{\varepsilon_0}}{2} [\vc{E}(\vc{r}) \otimes\uv{e}_\text{e} +  c\vc{B}(\vc{r})\otimes\uv{e}_\text{m}]\,,
% \end{equation}
where $\uv{e}_{\text{e}/\text{m}}$ are orthonormal vectors in a two-dimensional `electric-magnetic' vector space $\mathbb{C}^2$, while $\vc{E}(\vc{r})$ and $\vc{H}(\vc{r})$ are complex phasors living in a three-dimensional function-valued vector space %${\mathbb{C}^3\times L^2(\mathbb{R}^3)}$ 
and $\otimes$ is an outer product. This means that the bispinor lives in ${\mathbb{C}^2\times\mathbb{C}^3\times L^2(\mathbb{R}^3)}$, where the $\mathbb{C}^2$ subspace is the span of the $\uv{e}_\text{e}$ and $\uv{e}_\text{m}$ basis vectors, while the $\mathbb{C}^3$ subspace is the span of the three-dimensional vector orientations $\uv{x}$, $\uv{y}$, and $\uv{z}$. The two subspaces $\mathbb{C}^2$ and $\mathbb{C}^3$ are independent. For example, $\vc{\psi} = \uv{z}\otimes\uv{e}_{\text{m}}$ would represent a pure magnetic field, with unit energy density, pointing in the $\uv{z}$-direction, while a $\uv{x}$-polarised plane wave propagating along $z$ and with unit energy density is expressed as $\vc{\psi}(\vc{r}) = \frac{1}{\sqrt{2}}\left( \uv{x}\otimes\uv{e}_{\text{e}}+\uv{y}\otimes\uv{e}_{\text{m}}\right) e^{i k z}$.
In \cref{eq:bispinor} we made a particular choice of normalisation. The norm of our bispinor is taken to be the total energy density:\footnote{Another sensible choice for normalisation (especially in vacuum) is the number density of photons $W/(\hbar\omega)$, which gives bispinors dimensions of quantum wavefunction [\si{\meter^{-3/2}}] (see \cref{app:bispinor}).}
\begin{equation}
    \norm{\vc\psi}^2=\vc\psi^\dagger\vc\psi=\tfrac{1}{4}(\varepsilon \abs{\vc{E}}^2 + \mu \abs{\vc{H}}^2)=W(\vc{r})\,.
\end{equation}
The basis in the $\mathbb{C}^2$ space given by $\lbrace\uv{e}_{\text{e}},\uv{e}_{\text{m}}\rbrace$ is but one of many other possible choices. Notice that the $\mathbb{C}^2$ space is isomorphic to the space of Jones vectors that describe electric field polarisation in a plane wave, and as such one can define basis vectors in $\mathbb{C}^2$ that are inspired by the diagonal and circular bases often used in plane wave polarisation, which for this bispinor $\mathbb{C}^2$ space we termed parallel/antiparallel (meaning that ${\vc{E}\upharpoonleft \! \upharpoonright\vc{H}}$ or ${\vc{E}\upharpoonleft \! \downharpoonright\vc{H}}$) and right/left-handed (which are fields with either positive or negative helicity) in \cite{golat2024electromagnetic}
\begin{equation}
    \label{eq:epa_eRL_unit_vectors}
    \uv{e}_\text{p/a}=\tfrac{1}{\sqrt{2}}(\uv{e}_\text{e}\pm\uv{e}_\text{m}),\quad
    \uv{e}_\text{R/L}=\tfrac{1}{\sqrt{2}}(\uv{e}_\text{e}\mp\ii\uv{e}_\text{m})\,.
\end{equation}
A crucial realisation is that one can express any electromagnetic field as a sum of two components in any of the orthonormal bases:
\begin{equation}
    \vc\psi = \vc\psi_\text{e} +\vc\psi_\text{m} = \vc\psi_\text{a}+\vc\psi_\text{p}=\vc\psi_\text{R}+\vc\psi_\text{L}\,.
\end{equation}
These components are projections of the full electromagnetic field into each of the bases and can be mathematically obtained via an inner product $\vc\psi_i = (\uv{e}_i^\ast\vdot\vc\psi)\otimes\uv{e}_{i}= \vc{F}_i\otimes\uv{e}_{i}$ with $i\in\{\text{e,m,p,a},\textsc{r,l}\}$. The complex vector $\vc{F}_i$ is defined in $\mathbb{C}^3$; for instance, $\vc{F}_\text{e}=\frac{1}{2}\sqrt{\varepsilon}\vc{E}$ is proportional to the electric field vector (but normalised such that its norm squared is the electric field energy). Then one can define the energy density for each separate field projection. For example, $W_\text{e}$ is the energy density contained in the electric field `component' of an electromagnetic field, while $W_\text{R}$ is the energy density carried by the right-handed component. Mathematically, the different energy densities can be obtained as ${W_i=\abs{\vc{F}_i}^2}$ with $i\in\{\text{e,m,p,a},\textsc{r,l}\}$, which means (thanks to Parseval's theorem) that the total energy can be written as the sum of the energies of each of the two orthonormal projections in each basis
\begin{equation}
    W_0\equiv W= W_\text{e}+W_\text{m}= W_\text{a}+W_\text{p}=
    W_\text{R}+W_\text{L}\,.
\end{equation}
In analogy to Stokes parameters, one can also define quantities that are differences between the projections:
\begin{equation}
    \begin{alignedat}{2}
        W_1&=W_\text{e}&&-W_\text{m}\,,\\
        W_2&=W_\text{a}&&-W_\text{p}\,,\\
        W_3&=W_\text{R}&&-W_\text{L}\,.
    \end{alignedat}
\end{equation}
These energy densities are related to other electromagnetic scalar quantities. For example, in \citer{golat2024electromagnetic} we show that the electromagnetic `active' helicity density is $\mathfrak{S}=W_3/\omega$ \cite{Trueba1996,Afanasiev1996,Berry2019}, while $W_2/\omega$ is the reactive helicity density \cite{NietoVesperinas2021,kamenetskii2015microwave}; finally, $cW_1$ is the reactive power density (reactive power flow per unit area) \cite{NietoVesperinas2022, Jackson1998} and, similarly, $cW_0$ can be understood as the active power density. 

The same procedure can be performed for canonical momentum densities $\vc{p}$ and spin angular momentum densities $\vc{S}$. The canonical momentum for each projection $i\in\{\text{e,m,p,a},\textsc{r,l}\}$ can be expressed\footnote{We use the notation $\vc{a}^*\vdot(\grad)\vc{b} \equiv a_x^*  \grad b_x+a_y^*  \grad b_y+a_z^*  \grad b_z$ which can be used to obtain a weighted average of the phase gradients of each component $\Im[\vc{a}^*\vdot(\grad)\vc{a}] = |a_x|^2 \grad {\mathrm{Arg}} (a_x) + |a_y|^2 \grad {\mathrm{Arg}} (a_y) + |a_z|^2 \grad {\mathrm{Arg}} (a_z)$.} as ${\vc{p}_i=\tfrac{1}{\omega}\Im[\vc{F}_i^*\vdot(\grad)\vc{F}_i]}$, and the total canonical momentum is the sum of each pair of projections:
\begin{equation}
    \vc{p}_0\equiv \vc{p}= \vc{p}_\text{e}+\vc{p}_\text{m}= \vc{p}_\text{a}+\vc{p}_\text{p}=
    \vc{p}_\text{R}+\vc{p}_\text{L}\,.
\end{equation}
Similarly, the spin angular momentum density of each projection is ${\vc{S}_i=\tfrac{1}{\omega}\Im[\vc{F}_i^*\cp\vc{F}_i]}$, and the total is:
\begin{equation}
    \vc{S}_0\equiv \vc{S}= \vc{S}_\text{e}+\vc{S}_\text{m}= \vc{S}_\text{a}+\vc{S}_\text{p}=
    \vc{S}_\text{R}+\vc{S}_\text{L}\,.
\end{equation}
Having defined these observables for each projection, we can define their differences like we did in the case of $W_A$ where $A\in\{0,1,2,3\}$:
\begin{equation}
    \begin{alignedat}{4}
        \vc{p}_1&=\vc{p}_\text{e}&&-\vc{p}_\text{m}\,,&\qquad \vc{S}_1&=\vc{S}_\text{e}&&-\vc{S}_\text{m}\,,\\
        \vc{p}_2&=\vc{p}_\text{a}&&-\vc{p}_\text{p}\,,&\qquad 
        \vc{S}_2&=\vc{S}_\text{a}&&-\vc{S}_\text{p}\,,\\
        \vc{p}_3&=\vc{p}_\text{R}&&-\vc{p}_\text{L}\,,&\qquad 
        \vc{S}_3&=\vc{S}_\text{R}&&-\vc{S}_\text{L}\,.
    \end{alignedat}
\end{equation}
Many well-known quadratic quantities of electromagnetic fields are `hidden' inside these definitions as we highlighted in \citer{golat2024electromagnetic}. For instance, $\vc{S}_0$ represents the total electromagnetic spin angular momentum density, and $\vc{S}_1=\vc{S}_\text{e}-\vc{S}_\text{m}$ vanishes in paraxial light (much like other reactive quantities). The flow of active 
and reactive power can be represented by the real and imaginary parts of the complex Poynting vector $\vc{\Pi}=\tfrac{1}{2}\vc{E}\cp\vc{H}^*=\omega c(\vc{S}_3+\ii\vc{S}_2)$ \cite{kamenetskii2015microwave,Jackson1998}. The total canonical momentum density of light (also sometimes called orbital linear momentum) is $\vc{p}_0$, while $\vc{p}_3=\vc{p}_\text{R}-\vc{p}_\text{L}$ is the chiral momentum density \cite{Bliokh2014,Vernon2023} (the difference in momentum carried by the two light helicity components). The quantities $\vc{p}_1=\vc{p}_\text{e}-\vc{p}_\text{m}$ and $\vc{p}_2=\vc{p}_\text{a}-\vc{p}_\text{p}$ are the reactive achiral and reactive chiral momentum densities, respectively.
We can also think of these `observables' in terms of operators \cite{Bliokh2014}. Let's say we have a time averaged observable $A(\vc{r})$ which is quadratic in the fields; then we can see it as a local expectation value of an operator $\hat{A}$ acting on the wavefunction $\vc\psi(\vc{r})$, such that the total integrated expectation value is:
\begin{equation}
    \expval{\hat{A}}{\psi}=\iiint[\underbrace{\vc{\psi}^\dagger(\vc{r})\hat{A}\vc{\psi}(\vc{r})]}_{A(\vc{r})}\dd{V}
\end{equation}
where $\expval{\hat{A}}{\psi}$ is the time averaged integrated expectation value (e.g., the total energy of the field in a given integration region), $A(\vc{r})=\vc{\psi}^\dagger(\vc{r})\hat{A}\vc{\psi}(\vc{r})$ is the local density (e.g., energy density $W(\vc{r})=\vc{\psi}^\dagger(\vc{r})\hat{W}\vc{\psi}(\vc{r})$) and finally $\hat{A}$ is the operator associated with the observable (e.g., for energy density in our normalisation it will be the identity $\hat{W}=\hat{I}$, such that $W(\vc{r})=\vc{\psi}^\dagger(\vc{r})\vc{\psi}(\vc{r})$). The energy, momentum, and spin, in our convention, are related to the following operators (see \cref{app:operators}):
\begin{alignat}{2}
    W(\vc{r})&=\vc{\psi}^\dagger(\vc{r})\hat{W}\vc{\psi}(\vc{r})&&=\vc\psi^\dagger\hat{I}\vc\psi\,,\notag\\
    \vc{p}(\vc{r})&=\vc{\psi}^\dagger(\vc{r})\,\uv{p}\,\vc{\psi}(\vc{r})&&=\tfrac{1}{\omega}\Im[\vc\psi^\dagger\vdot(\grad)\vc\psi]\,,\\
    \vc{S}(\vc{r})&=\vc{\psi}^\dagger(\vc{r})\,\uv{S}\,\vc{\psi}(\vc{r})&&=\tfrac{1}{\omega}\Im(\vc\psi^\dagger\cp\vc\psi)\,.\notag
\end{alignat}
Operators for quantities such as $W_A$, $\vc{p}_A$, and $\vc{S}_A$ can then be obtained by multiplying by an appropriate Pauli matrix, for example $\hat{W}_A=\hat{W}\hat{P}_A$ (see \cref{tab:transposed_matrix_basis}).
\begin{table}[th]
\centering
\renewcommand{\arraystretch}{1.2}
\caption{Changing the basis of linear transformations in $\mathbb{C}^2$, with Pauli matrices: $\tens{\sigma}_1=(\smqty{\pmat{1}})$, $\tens{\sigma}_2=(\smqty{0&-\ii\\\ii&0})$, $\tens{\sigma}_3=(\smqty{\pmat{3}})$.}\vspace{.8em}
\begin{tabular}{|c||c|c|c|c|}
\hhline{|-||----|}
$\mathbb{C}^2$ basis & $\hat{P}_0$ & $\hat{P}_1$ & $\hat{P}_2$ & $\hat{P}_3$ \\ \hhline{:=::====:}
$\lbrace\uv{e}_{\text{R}},\uv{e}_{\text{L}}\rbrace$ & $\tens{I}$ & $\tens{\sigma}_1$ & $\tens{\sigma}_2$ & $\tens{\sigma}_3$ \\ 
\hhline{|-||----|}
$\lbrace\uv{e}_{\text{e}},\uv{e}_{\text{m}}\rbrace$ & $\tens{I}$ & $\tens{\sigma}_3$ & $-\tens{\sigma}_1$ & $-\tens{\sigma}_2$ \\ 
\hhline{|-||----|}
$\lbrace\uv{e}_{\text{p}},\uv{e}_{\text{a}}\rbrace$ & $\tens{I}$ & $\tens{\sigma}_1$ & $-\tens{\sigma}_3$ & $\tens{\sigma}_2$ \\ 
\hhline{|-||----|}
\end{tabular}
\label{tab:transposed_matrix_basis}
\end{table}

In the bispinor formalism \cite{Bliokh2014,golat2024electromagnetic}, the expressions for the rate of transfer of energy (power absorption), helicity, linear momentum (optical force) and angular momentum (optical torque) to an electromagnetic dipole take a very simple form (see \cref{app:operatorPHFG}):
\begin{alignat}{3}
    \langle\mathcal{P}_\text{abs}\rangle&=2k\,\Im(\vc{\psi}^\dagger c\hat{W}_0\vc{\pi})&&-\tfrac{k^4}{3\pi}(\vc{\pi}^\dagger c\hat{W}_0\vc{\pi})\,&&,\label{eq:power}
\\
\omega\langle\dot{\mathcal{H}}_\text{abs}\rangle&=\underbrace{2k\,\Im(\vc{\psi}^\dagger c\hat{W}_3\vc{\pi})}_\text{extinction}&&-\underbrace{\tfrac{k^4}{3\pi}(\vc{\pi}^\dagger c\hat{W}_3\vc{\pi})}_\text{scattering}\,&&,\label{eq:helicity}
\\\langle{\vc{F}}\rangle&=2k\,\Im(\vc{\pi}^\dagger c\uv{p}_0\,\vc{\psi})&&-\tfrac{k^5}{6\pi }(\vc{\pi}^\dagger c\uv{S}_3\vc{\pi})\,&&,\label{eq:force}
\\
    \langle{\vc{\varGamma}}\rangle&=\underbrace{2k\,\Im(\vc{\pi}^\dagger c\uv{S}_0\,\vc{\psi})}_\text{interaction}&&-\underbrace{\tfrac{k^4}{3\pi }(\vc{\pi}^\dagger c\uv{S}_0\vc{\pi})}_\text{recoil}\,&&,\label{eq:torque}
\end{alignat}
where we defined the electromagnetic dipole bispinor \cite{Bliokh2014}:
% \hbadness 1947\relax
\begin{equation}
    \vc{\pi}(\vc{r})=\tfrac{1}{2} (\vc{p}(\vc{r})/\sqrt{\varepsilon}\otimes\uv{e}_\text{e} +  \sqrt{\mu} \vc{m}(\vc{r})\otimes\uv{e}_\text{m})\,.
\end{equation}

The first term (extinction or interaction term) in \cref{eq:power,eq:force,eq:torque} has been previously presented in \citer{Bliokh2014}, while the rest are newly derived here providing a complete picture. Notice that the expression for the absorbed helicity is very intuitive since the helicity density is $\mathfrak{S}=W_3/\omega$.
We can also see that the interaction term of the force depends on the canonical momentum $\vc{p}_0$, but the recoil depends on the kinetic momentum $k\vc{S}_3=\Re\vc{\Pi}/c^2$. These expressions can be further simplified by assuming linearity,
\begin{equation}
\vc{\pi}=\tens{A}\vc{\psi},
\end{equation}
where $\tens{A}$ can be represented in the usual electromagnetic basis by the matrix in \cref{eq:generaltensorpolarisability}, however, as we showed in \citers{golat2024electromagnetic,Golat2023} it is more convenient to express complex polarisabilities in a basis independent way via the use of Stokes-parameter-inspired polarisabilities $\alpha_A$ with $A=0,1,2,3$ (see \cref{tab:polarisabilities}). This comes from the fact that the linear dipolar response can be decomposed into a basis spanned by Pauli matrices as follows:\footnote{Note that we have introduced an additional factor of 1/2 so that the polarisabilities correspond to the sums and differences of the individual projections. This differs from the convention used in \cite{golat2024electromagnetic}, where they were defined as half the sums and differences.}
\begin{equation}
    \tens{A} = \frac{1}{2}\sum_{A=0}^3 \alpha_A \hat{P}_A\,,
\end{equation}
where this expansion is valid for all bases, one just has to take the appropriate Pauli matrix from \cref{tab:transposed_matrix_basis}.
\begin{table}[ht!]
\centering
\caption{Definition of complex polarisabilities written as sum/difference of polarisabilities of the projections into the different basis unit vectors, and how they are conventionally called in the literature.}
\vspace{.8em}
\renewcommand{\arraystretch}{1.3} % Default value: 1
\begin{tabular}{|c||c|c|}
\hhline{|-||--|}
 Polarisability& As a sum/difference\, &Conventional\\
\hhline{:=::==:}
$\alpha_0$&\multicolumn{2}{c|}{$\alpha_\text{R}+\alpha_\text{L}=\alpha_\text{a}+\alpha_\text{p}=\alpha_\text{e}+\alpha_\text{m}$}\\
\hhline{|-||--|}
$\alpha_1$&\multicolumn{2}{c|}{$\,\alpha_\text{e}-\alpha_\text{m}$}\\
\hhline{|-||--|}
$\alpha_2$&$\alpha_\text{a}-\alpha_\text{p}$& $\mathllap{-\,}2\alpha_\text{t}$
\\
\hhline{|-||--|}
$\alpha_3$&$\alpha_\text{R}-\alpha_\text{L}$& $2\alpha_\text{c}$
\\
\hhline{|-||--|}
\end{tabular}
    \label{tab:polarisabilities}
\end{table}%

\begin{table*}[ht!]
\centering
\caption{Cross sections involved in bi-isotropic dipolar interactions expressed in terms of complex polarisabilities.}
\vspace{1em}
\renewcommand{\arraystretch}{1.6}
\begin{tabular}{|cc||c|cc||c|cc|}
\hhline{|--||-|--||-|--|}
\multicolumn{2}{|c||}{\multirow{2}{*}{\parbox[c]{1cm}{\centering}}} &
  \multirow{2}{*}{\parbox[c]{0.5cm}{\centering $\sigma_A$}} &
  \multicolumn{2}{c||}{\centering Power cross section} &
  \multirow{2}{*}{\parbox[c]{0.5cm}{\centering $\gamma_A$}} &
  \multicolumn{2}{c|}{\centering \hspace{6pt}Helicity rate cross section\hspace{6pt}} \\ 
\hhline{|~~||~|--||~|--|}
\multicolumn{2}{|c||}{} & &
  \multicolumn{1}{c|}{\hspace*{2em}\cellcolor{active}\phantom{r}active\phantom{e}\hspace*{2em}} &
  \multicolumn{1}{c||}{\hspace*{2em}\cellcolor{reactive}reactive\hspace*{2em}} &&
  \multicolumn{1}{c|}{\hspace*{2em}\cellcolor{active}\phantom{r}active\phantom{e}\hspace*{2em}} &
  \multicolumn{1}{c|}{\hspace*{2em}\cellcolor{reactive}reactive\hspace*{2em}} 
  \\ \hhline{:==::=:==::=:==:}
\multicolumn{1}{|c|}{\multirow{4}{*}{\parbox[c]{12pt}{\centering\rotatebox[origin=r]{90}{Extinction}}}} &
  \multirow{2}{*}{\rotatebox[origin=r]{90}{achiral}} & $\sigma_\text{ext}^0$ &
  \multicolumn{2}{c||}{\cellcolor{active}$k\Im(\alpha_0)$} &
  $\gamma_\text{ext}^3$ &
  \multicolumn{2}{c|}{\cellcolor{active}$k\Im(\alpha_0)$} 
  \\ 
\hhline{|~~||-|--||-|--|}
\multicolumn{1}{|c|}{} &
   & $\sigma_\text{ext}^1$ &
   \multicolumn{2}{c||}{\cellcolor{reactive}$k\Im(\alpha_1)$} &
   $\gamma_\text{ext}^2$ &
   \multicolumn{2}{c|}{\cellcolor{reactive}$k\Re(\alpha_1)$} 
  \\ 
\hhline{|~-||-|--||-|--|} 
\multicolumn{1}{|c|}{} &
  \multirow{2}{*}{\rotatebox[origin=r]{90}{chiral}} & $\sigma_\text{ext}^2$ &
  \multicolumn{2}{c||}{\cellcolor{reactive}$k\Im(\alpha_2)$} &
  $\gamma_\text{ext}^1$ &
  \multicolumn{2}{c|}{\cellcolor{reactive}$\mathllap{-\,}k\Re(\alpha_2)$} 
  \\ 
\hhline{|~~||-|--||-|--|}
\multicolumn{1}{|c|}{} &
   & $\sigma_\text{ext}^3$ &
   \multicolumn{2}{c||}{\cellcolor{active}$k\Im(\alpha_3)$} &
   $\gamma_\text{ext}^0$ &
   \multicolumn{2}{c|}{\cellcolor{active}$k\Im(\alpha_3)$} 
  \\ \hhline{:==::=:==::=:==:}
\multicolumn{1}{|c|}{\multirow{4}{*}{\parbox[c]{12pt}{\centering\rotatebox[origin=r]{90}{Scattering}}}} &
  \multirow{2}{*}{\rotatebox[origin=r]{90}{achiral}} & $\sigma_\text{sca}^0$ &
  \multicolumn{2}{c||}{\cellcolor{active}$\tfrac{k^4}{12\pi}(\abs{\alpha_0}^2+\abs{\alpha_1}^2+\abs{\alpha_2}^2+\abs{\alpha_3}^2)$} &
  $\gamma_\text{sca}^3$ &
  \multicolumn{2}{c|}{\cellcolor{active}$\tfrac{k^4}{12\pi}(\abs{\alpha_0}^2-\abs{\alpha_1}^2-\abs{\alpha_2}^2+\abs{\alpha_3}^2)$} 
  \\  
\hhline{|~~||-|--||-|--|} 
\multicolumn{1}{|c|}{} &
   &$\sigma_\text{sca}^1$ &
   \multicolumn{2}{c||}{\cellcolor{reactive}$\tfrac{k^4}{6\pi}[\Re(\alpha_0^\ast\alpha_1)-\Im(\alpha_2^\ast\alpha_3)]$} &
   $\gamma_\text{sca}^2$ &
   \multicolumn{2}{c|}{\cellcolor{reactive}$\tfrac{k^4}{6\pi}[\Re(\alpha_3^\ast\alpha_2)+\Im(\alpha_1^\ast\alpha_0)]$} 
  \\ 
\hhline{|~-||-|--||-|--|} 
\multicolumn{1}{|c|}{} &
  \multirow{2}{*}{\rotatebox[origin=r]{90}{chiral}} & $\sigma_\text{sca}^2$ &
  \multicolumn{2}{c||}{\cellcolor{reactive}$\tfrac{k^4}{6\pi}[\Re(\alpha_0^\ast\alpha_2)-\Im(\alpha_3^\ast\alpha_1)]$} &
  $\gamma_\text{sca}^1$ &
  \multicolumn{2}{c|}{\cellcolor{reactive}$\tfrac{k^4}{6\pi}[\Re(\alpha_3^\ast\alpha_1)+\Im(\alpha_0^\ast\alpha_2)]$} 
  \\ 
\hhline{|~~||-|--||-|--|} 
\multicolumn{1}{|c|}{} & 
   & $\sigma_\text{sca}^3$ &
   \multicolumn{2}{c||}{\cellcolor{active}$\tfrac{k^4}{6\pi}[\Re(\alpha_0^\ast\alpha_3)-\Im(\alpha_1^\ast\alpha_2)]$} &
   $\gamma_\text{sca}^0$ &
   \multicolumn{2}{c|}{\cellcolor{active}$\tfrac{k^4}{6\pi}[\Re(\alpha_3^\ast\alpha_0)+\Im(\alpha_1^\ast\alpha_2)]$}  
  \\ \hhline{:==::=:==::=:==:}
\multicolumn{1}{|c|}{\multirow{4}{*}{\parbox[c]{12pt}{\centering\rotatebox[origin=r]{90}{Absorption}}}} &
  \multirow{2}{*}{\rotatebox[origin=r]{90}{achiral}} & $\sigma_\text{abs}^0$ &
  \multicolumn{2}{c||}{\cellcolor{active}$\sigma_\text{ext}^0-\sigma_\text{sca}^0$} &
  $\gamma_\text{abs}^3$ &
  \multicolumn{2}{c|}{\cellcolor{active}$\gamma_\text{ext}^3-\gamma_\text{sca}^3$} 
  \\  
\hhline{|~~||-|--||-|--|} 
\multicolumn{1}{|c|}{} &
   & $\sigma_\text{abs}^1$ &
  \multicolumn{2}{c||}{\cellcolor{reactive}$\sigma_\text{ext}^1-\sigma_\text{sca}^1$} &
  $\gamma_\text{abs}^2$ &
  \multicolumn{2}{c|}{\cellcolor{reactive}$\gamma_\text{ext}^2-\gamma_\text{sca}^2$} 
  \\ 
\hhline{|~-||-|--||-|--|} 
\multicolumn{1}{|c|}{} &
  \multirow{2}{*}{\rotatebox[origin=r]{90}{chiral}} & $\sigma_\text{abs}^2$ &
  \multicolumn{2}{c||}{\cellcolor{reactive}$\sigma_\text{ext}^2-\sigma_\text{sca}^2$} &
  $\gamma_\text{abs}^1$ &
  \multicolumn{2}{c|}{\cellcolor{reactive}$\gamma_\text{ext}^1-\gamma_\text{sca}^1$} 
  \\ 
\hhline{|~~||-|--||-|--|} 
\multicolumn{1}{|c|}{} & 
   & $\sigma_\text{abs}^3$ &
   \multicolumn{2}{c||}{\cellcolor{active}$\sigma_\text{ext}^3-\sigma_\text{sca}^3$} &
   $\gamma_\text{abs}^0$ &
   \multicolumn{2}{c|}{\cellcolor{active}$\gamma_\text{ext}^0-\gamma_\text{sca}^0$}  
  \\ \hhline{|--||-|--||-|--|}
\end{tabular}
\label{tab:cross sections}
\end{table*}

\section{Bi-isotropic linear interactions}
The response of a bi-isotropic particle has no preferred spatial direction, so \cref{eq:power,eq:helicity,eq:force,eq:torque} can be expressed entirely in terms of observables associated with the incident fields, namely the energy densities \( W_A \), the momentum densities \( \vc{p}_A \), and the spin densities \( \vc{S}_A \), multiplied by coefficients that depend on the polarisabilities. The absorbed power \cref{eq:power} and its individual terms---the extinguished or scattered power---can be written in the following form (see \cref{app:power}):
\begin{equation}
    \langle\mathcal{P}\rangle=c(\underbrace{\sigma_0 W_0}_\text{active}+\underbrace{\sigma_1 W_1}_\text{reactive}+\overbrace{\underbrace{\sigma_2 W_2}_\text{reactive}+\underbrace{\sigma_3 W_3}_\text{active}}^\text{chiral})\,,\label{eq:dipolar_power}
\end{equation} 
where both $\langle\mathcal{P}\rangle$ and the cross-sections $\sigma_A$ will have a subscript depending on whether we mean absorption, extinction, or scattering. Each term in the sum over \(A \in \{0,1,2,3\}\) can be interpreted as either \emph{active} (\(A \in \{0,3\}\)), which can be measured using paraxial/far-field illumination, or \emph{reactive} (\(A \in \{1,2\}\)), which cannot (because $W_1=W_2=0$ in far-field paraxial light). Each term can be classified as \emph{chiral} or \emph{achiral}, depending on its behaviour under parity. The corresponding cross-sections $\sigma_A$ are listed in \cref{tab:cross sections}. If the illumination is a linearly polarised plane wave (or paraxial beam), we will have $W_1=W_2=W_3=0$ and this expression becomes the well known relationship between power $\langle\mathcal{P}\rangle$, field intensity $I_0=cW_0$ ($\si{\watt\per\meter\squared}$) and the cross section $\sigma_0$ ($\si{\meter\squared}$): $\langle\mathcal{P}\rangle=\sigma_0I_0$. If the illumination remains paraxial but an arbitrary polarisation state is allowed---including circular and elliptical polarisations---then \( W_1 = W_2 = 0 \) as in all far-field paraxial light, while \( W_0 = W_\text{R} + W_\text{L} \) and \( W_3 = W_\text{R} - W_\text{L} \). In that case, the power can be written in the following form:
\begin{equation}\label{eq:power_paraxial}
    \langle\mathcal{P}\rangle=\underbrace{(\sigma_0+\sigma_3)}_{\sigma_\text{R}}cW_\text{R}+\underbrace{(\sigma_0-\sigma_3)}_{\sigma_\text{L}}cW_\text{L}\,.
\end{equation}
This formulation makes it evident that the `active chiral' cross section $\sigma_3=(\langle\mathcal{P}_\text{R}\rangle-\langle\mathcal{P}_\text{L}\rangle)/(2cW_0)$ quantifies the differential response to right- and left-handed light---an effect known as \emph{circular dichroism}.\footnote{Traditionally reported as an ellipticity angle or a dissymmetry factor
\(g=2({\langle\mathcal{P}_\text{ext}^\text{R}\rangle-\langle\mathcal{P}_\text{ext}^\text{L}}\rangle)/{(\langle\mathcal{P}_\text{ext}^\text{R}\rangle+\langle\mathcal{P}_\text{ext}^\text{L}\rangle)}=2{\sigma_\text{ext}^3}/{\sigma_\text{ext}^0}\,.\)
} Since the reactive quantities \(W_1\) and \(W_2\) are nonzero only in near fields or structured beams, reactive cross sections \(\sigma_1\) and \(\sigma_2\) cannot be measured using paraxial light.
They may be interpreted analogously as differences in electric and magnetic responses, $\sigma_1\propto\langle\mathcal{P}_\text{e}\rangle-\langle\mathcal{P}_\text{m}\rangle$, and between parallel and antiparallel responses,
$\sigma_2\propto\langle\mathcal{P}_\text{a}\rangle-\langle\mathcal{P}_\text{p}\rangle$. To access these, one requires standing or evanescent waves (see \cref{app:measuring}), which feature regions dominated by purely electric, magnetic, parallel, or antiparallel fields.
Notably, the reactive chiral response \(\sigma_2\)---also known as \emph{reactive dichroism}~\cite{NietoVesperinas2021} or \emph{trochroidal dichroism}~\cite{McCarthy2020Jul}---has already been experimentally observed in the scattered power under near field illumination reported in \citer{McCarthy2020Jul}.

The scattered power is never negative $\langle\mathcal{P}_\text{sca}\rangle\geq0$ and since this is true for any illumination, it means that the scattering cross section operator is positive semidefinite, leading to the following cross section inequality (see \cref{app:power}):
\begin{equation}\label{eq:ineq}
    {\sigma}_\text{sca}^0\geq\sqrt{({\sigma}_\text{sca}^1)^2+({\sigma}_\text{sca}^2)^2+({\sigma}_\text{sca}^3)^2}\,,
\end{equation}
which suggests that the particle scattering response has its own symmetry sphere much like the electromagnetic field itself \cite{golat2024electromagnetic}. In addition, a \emph{passive} particle (no gain) must satisfy the condition $\langle\mathcal{P}_\text{abs}\rangle\geq0$, which together with $\langle\mathcal{P}_\text{ext}\rangle=\langle\mathcal{P}_\text{abs}\rangle+\langle\mathcal{P}_\text{sca}\rangle$ leads to the inequality $\langle\mathcal{P}_\text{ext}\rangle\geq\langle\mathcal{P}_\text{sca}\rangle\geq0$ equivalent to $\sigma^0_\text{ext}\geq\sigma^0_\text{sca}$ for arbitrary illumination. Following the same reasoning as for scattering cross sections, we get the same inequality (\cref{eq:ineq}) but for both absorption and extinction cross sections in passive particles. The extinction cross section inequality for a passive particle can be rewritten (using $\sigma^A_\text{ext}=k\Im(\alpha_A)$ from \cref{tab:cross sections}) as a relationship between imaginary parts of polarisabilities:
\begin{equation}\label{eq:extinction_inequality}
    \Im{\alpha_0}\geq\sqrt{(\Im{\alpha_1})^2+(\Im{\alpha_2})^2+(\Im{\alpha_3})^2}\,.
\end{equation}
This constraint illustrates a well known fact that particles with chiral extinction $\Im{\alpha_3}\neq0$ must have non-zero magnetic response $\Im{\alpha_\text{m}}=\Im{(\alpha_0-\alpha_1)}/2\neq0$. Since the majority of particles are reciprocal ($\alpha_2=0$) we can rewrite \cref{eq:extinction_inequality} in that particular case as a lower (if $\Im{\alpha_\text{e}}>0$) or upper (if $\Im{\alpha_\text{e}}<0$) bound for the magnetic response:
\begin{equation}
    \Im{\alpha_\text{e}}\Im{\alpha_\text{m}}\geq(\Im{\alpha_\text{c}})^2\,.
\end{equation}
This special case of reciprocal particles has been previously derived and experimentally tested for split ring arrays \cite{Sersic2012May}, while \cref{eq:extinction_inequality} remains most general.

The helicity rates \cref{eq:helicity} will have a similar form to the power:
\begin{equation}
    \omega\langle\dot{\mathcal{H}}\rangle=c(\underbrace{\gamma_3 W_3}_\text{active}+\underbrace{\gamma_2 W_2}_\text{reactive}+\overbrace{\underbrace{\gamma_1 W_1}_\text{reactive}+\underbrace{\gamma_0 W_0}_\text{active}}^\text{chiral})\label{eq:dipolar_helicity}\,,
\end{equation} 
where again both helicity rates $\langle\dot{\mathcal{H}}\rangle$ and helicity cross sections $\gamma_A$ will have a subscript depending on whether we mean absorption, extinction, or scattering. Notice that since helicity is odd under the parity transformation, the numbering of chiral and achiral terms has switched places. We can again look at the paraxial limit with \( W_0 = W_\text{R} + W_\text{L} \) and \( W_3 = W_\text{R} - W_\text{L} \) and \( W_1 = W_2 = 0 \):
\begin{equation}
    \omega\langle\dot{\mathcal{H}}\rangle=\underbrace{(\gamma_0+\gamma_3)}_{\gamma_\text{R}}cW_\text{R}+\underbrace{(\gamma_0-\gamma_3)}_{\gamma_\text{L}}cW_\text{L}\,.
\end{equation}
This looks very intuitive and looking at \cref{tab:cross sections} we see that
the rate at which the energy and helicity of a given handedness are extinguished is equal ($\sigma_\text{ext}^\text{R/L}=\pm\gamma_\text{ext}^\text{R/L}$) or in terms of cross sections $\sigma_\text{ext}^{0/3}=\gamma_\text{ext}^{3/0}$. However, notice that the same is not true for absorption or scattering unless the particle is dual symmetric ($\alpha_1=\alpha_2=0$). As a consequence, for particles breaking duality, helicity cross sections $\gamma_A$ are needed to provide a complete description, as they cannot be written just in terms of power cross sections $\sigma_A$.

Intriguingly, these very same cross sections ($\sigma_A$ and $\gamma_A$) also appear in the complete expression for the dipolar force (valid for chiral and nonreciprocal particles):
\begin{align}\label{eq:dipolar_force}
    \langle\vc{F}\rangle=\sum_{A=0}^3(\underbrace{\vphantom{\gamma_\text{sca}^A}\tfrac{1}{2}\Re\alpha_A\grad{W_A}}_\text{gradient}\underbrace{+\vphantom{\tfrac{1}{2}}\sigma_\text{ext}^Ac\vc{p}_A}_\text{pressure}\underbrace{-\tfrac{1}{2}\gamma_\text{sca}^Ack\vc{S}_A}_\text{recoil})\,,
\end{align}
suggesting the origin of each term.
The first group of terms corresponds to a conservative gradient force, $\langle\vc{F}_\text{gra}\rangle=-\grad{\langle U\rangle}$. It originates from the spatial variations of the induced dipole potential energy:
\begin{equation}
    \langle U\rangle=-\frac{1}{4}\Re(\vc{E}^\ast\!\vdot\vc{p}+\mu\vc{H}^\ast\!\vdot\vc{m})=-\frac{1}{2}\sum_{A=0}^3\Re\alpha_A {W_A}\,,
\end{equation}
which has again active, reactive, chiral, and achiral components.
This force attracts the particle to the regions where the potential $\langle U\rangle$ is lowest. The second group of terms is known as the radiation pressure force.\footnote{Notice that $c\vc{p}_A$ has units of pressure.} It is caused by the transfer of canonical linear momentum $\vc{p}_A$ carried by photons to the particle either through absorption ($\sigma_\text{abs}^A$) or scattering ($\sigma_\text{sca}^A$) which is why these terms depend on the power extinction cross sections $\sigma_\text{ext}^A=\sigma_\text{abs}^A+\sigma_\text{sca}^A$:
\begin{align}\label{eq:F_pres}
    \langle\vc{F}_\text{pre}\rangle=c(\underbrace{\sigma_\text{ext}^0 \vc{p}_0}_\text{active}+\underbrace{\sigma_\text{ext}^1 \vc{p}_1}_\text{reactive}+\overbrace{\underbrace{\sigma_\text{ext}^2 \vc{p}_2}_\text{reactive}+\underbrace{\sigma_\text{ext}^3 \vc{p}_3}_\text{active}}^\text{chiral})\,.
\end{align}
Each of these terms brings much intuition and relates the absorption of linear momentum to the breaking of symmetries by the dipolar particle: for instance, most matter reacts more strongly to electric than magnetic fields, breaking the electric-magnetic symmetry. This results in a pressure force $c \sigma_\text{ext}^1 \vc{p}_1 =\tfrac{1}{2} c(\sigma_\text{ext}^\text{e}-\sigma_\text{ext}^\text{m})(\vc{p}_\text{e}-\vc{p}_\text{m})$, proportional to the difference in the particle's extinction between electric and magnetic fields, multiplied by the vector difference in the electromagnetic momentum between those two components of the applied light. A similar argument applies to a chiral particle that breaks parity symmetry and therefore responds differently to right-handed and left-handed light, which then experiences a chiral pressure force $c \sigma_\text{ext}^3 \vc{p}_3 = \tfrac{1}{2}c(\sigma_\text{ext}^\text{R}-\sigma_\text{ext}^\text{L})(\vc{p}_\text{R}-\vc{p}_\text{L})$. Importantly, we want to stress that \cref{eq:F_pres} is an \textit{exact} equation valid in any arbitrary light field, including near fields and non-paraxial fields. The only assumptions are that the particle response is linear and bi-isotropic, and that the medium can be characterised by real valued $\varepsilon$ and $\mu$. The four terms are strictly nonconservative, as can be seen from ${\div\vc{p}_A=0}$, which suggests that their Helmholtz decomposition will involve only a solenoidal (`curl') term. Finally, we have the recoil terms, which are attributed to the non-isotropic radiation from the particle due to the coherent superposition of the electric and magnetic dipoles \cite{NietoVesperinas2010}. These very same anisotropies will also affect the scattered helicity \cref{eq:dipolar_helicity}, as they depend, like the recoil force, on the interaction between the electric and magnetic dipoles. Looking at \cref{eq:helicity,eq:force}, it follows that the recoil force depends on half the helicity scattering cross sections ($\gamma_\text{sca}^A/2$) and points against the direction of spin angular momentum densities as follows:
\begin{align}
    \!\!\!\langle\vc{F}_\text{rec}\rangle=-\frac{\omega}{2}(\underbrace{\gamma_\text{sca}^3 \vc{S}_3}_\text{active}+\underbrace{\gamma_\text{sca}^2 \vc{S}_2}_\text{reactive}+\overbrace{\underbrace{\gamma_\text{sca}^1 \vc{S}_1}_\text{reactive}+\underbrace{\gamma_\text{sca}^0 \vc{S}_0}_\text{active}}^\text{chiral})\,.\!\!\!
\end{align}
The coefficients in front of the recoil terms have been calculated before \cite{Golat2023}, but to our best knowledge, it has not been pointed out before that they are related to the helicity scattering cross sections. This, of course, means that a particle that does not scatter helicity also does not experience a recoil force.
Active recoil force terms are nonconservative, just like the radiation pressure force, because ${\div\vc{S}_{0/3}=0}$. While reactive terms have ${\div\vc{S}_{1/2}=\pm\tfrac{2}{c}W_{2/1}}$ (see \cite{golat2024electromagnetic}), which means that a Helmholtz decomposition of the force will include both conservative gradient and non-conservative curl terms. 

In the paraxial limit (reactive quantities with $A=1,2$ go to zero), the total force can be written as follows:
\begin{equation}
\begin{gathered}
    \!\!\!\!\!\!\langle\vc{F}\rangle=\tfrac{1}{2}\grad\Re(\alpha_\text{R}W_\text{R}+\alpha_\text{L}W_\text{L})+c(\sigma_\text{ext}^\text{R}\vc{p}_\text{R}+\sigma_\text{ext}^\text{L}\vc{p}_\text{L})\!\!\!\!\!\!\\
    -\tfrac{\omega}{2}(\gamma_\text{sca}^\text{R}\vc{S}_\text{R}+\gamma_\text{sca}^\text{L}\vc{S}_\text{L})\,,
\end{gathered}
\end{equation}
which means that one could predict the force on a dipole in a paraxial field simply by extracting $\sigma_\text{ext}^\text{R/L}$ and $\gamma_\text{sca}^\text{R/L}$ from a measurement of power extinction $\langle\mathcal{P}_\text{ext}\rangle$ and the helicity scattering rate $\langle\dot{\mathcal{H}}_\text{sca}\rangle$ when illuminated by circularly polarised light.

Finally, we write simplified expressions for the torque. The torque always points along the spin angular momentum densities; the interaction term depends on power extinction, while the recoil depends on scattering, making the \textit{total} torque proportional to \textit{absorption} cross sections
\begin{align}
    \langle\vc{\varGamma}\rangle=c(\underbrace{\sigma_\text{abs}^0 \vc{S}_0}_\text{active}+\underbrace{\sigma_\text{abs}^1 \vc{S}_1}_\text{reactive}+\overbrace{\underbrace{\sigma_\text{abs}^2 \vc{S}_2}_\text{reactive}+\underbrace{\sigma_\text{abs}^3 \vc{S}_3}_\text{active}}^\text{chiral})\,.
\end{align}
Looking back at \cref{eq:torque}, we can see that the torque has interaction (extinction) and recoil (scattering) parts, much like power or force. This is reflected in the cross sections as $\sigma_\text{abs}^A=\sigma_\text{ext}^A-\sigma_\text{sca}^A$. Consider the example of a perfectly absorbing particle ($\langle\mathcal{P}_\text{sca}\rangle=0$). Such particle will acquire power $\langle\mathcal{P}_\text{abs}\rangle$ from the energy densities $W_A$ lost by the field, it will experience force from absorbing the field's momenta $\vc{p}_A$, and torque from the absorbed spins $\vc{S}_A$. However, when we consider an absorbing and scattering particle, the recoil force will start to depend on $\vc{S}_A$ also via the helicity scattering cross sections, while the total torque will still depend on the absorbed spins alone. 
In the paraxial limit, the torque is given by:
\begin{align}
    \langle\vc{\varGamma}\rangle=c(\sigma_\text{abs}^\text{R}\vc{S}_\text{R}+\sigma_\text{abs}^\text{L}\vc{S}_\text{L})\,.
\end{align}

Light carries both orbital and spin angular momentum. From the above equations, it may seem that, when light is absorbed, only the spin angular momentum part of the total angular momentum is transferred to the dipole (via a torque). However, this is not true: the dipolar force \cref{eq:dipolar_force} may also make the dipole rotate about a specific point or axis. One can calculate the moment of the force about an origin (whose position is arbitrary) as ${\vc{M}=\vc{r}\cp\vc{F}}$ to discover that the orbital angular momentum is transferred via the action of the pressure force
\begin{equation}
    \begin{split}
    \!\!\!\langle\vc{M}\rangle=\!\sum_{A=0}^3[{}&\sigma_\text{ext}^Ac\vc{L}_A\!+\tfrac{1}{2}\vc{r}\!\cp\!(\Re\alpha_A\grad{W_A}\!-\gamma_\text{sca}^A\omega\vc{S}_A)]\,,\!\!\!
\end{split}
\end{equation}
where ${\vc{L}_A=\vc{r}\cp\vc{p}_A}$ are sum and differences in orbital angular momentum density of the illuminating light for $A\in\{0,1,2,3\}$. 

We wish to highlight that the elegance and conciseness of all the above expressions is in very stark contrast to the same quantities expressed in the usual electric and magnetic basis, which are long expressions featuring many terms---even when, as is usually done, the nonreciprocity of the particle is neglected. The above simple expressions work for an arbitrary dipolar particle with any linear response, including chirality and nonreciprocity.

\section{Conclusions}

Our results suggest that dipolar light–matter interactions are best understood through the lens of symmetry breaking. Expressing the full set of absorbed and scattered power and helicity rates, as well as total force and torque on a dipolar particle, using the different electric-magnetic bases for the field properties and the particle's response, clearly exposes the underlying symmetry principles and yields remarkably compact expressions. Within this framework, we have derived general inequalities that constrain the response of dipolar particles, with stricter bounds for passive systems. The formalism also reveals unexpected results—most notably, the appearance of helicity cross sections multiplying the spin densities in the recoil force, and the fact that quantities with $A=2$ (such as $\sigma_\text{sca}^2$ and $\gamma_\text{sca}^2$), typically associated with nonreciprocity, can remain nonzero even in reciprocal particles ($\alpha_2=0$) provided both electric–magnetic ($\alpha_1\neq0$) and right–left ($\alpha_3\neq0$) symmetries are broken simultaneously, suggesting a form of `false reciprocity breaking' in scattering (but not in extinction). Although our analysis focuses on dipolar responses, the same symmetry-based structure extends naturally to higher-order multipoles: indeed, \citer{Yuan2025Dec} recently showed that forces and torques on \emph{arbitrary multipoles} follow the same symmetry-governed expressions presented here for dipoles, suggesting that the power and helicity absorption and scattering in general light–matter interactions will also always obey these symmetry-based formulations.

\begin{acknowledgments}
% \paragraph{Acknowledgments}\!\!\!---\,\,\,%
%AJV acknowledges support from the EPSRC Grant EP/R513064/1,
SG and FJRF are supported by the EIC-Pathfinder-CHIRALFORCE (101046961) which is funded by Innovate UK Horizon Europe Guarantee (UKRI project 10045438). 
\end{acknowledgments}

% \hbadness 10000\relax
\bibliography{main}

\onecolumngrid
% \newpage

\maketitle
\section*{Supplementary Information}
\appendix
\crefalias{section}{appendix}    % tell cleveref: sections here are appendices
% (optional, if you reference subsections too)
\crefalias{subsection}{appendix}
\section{Bispinor operator formalism for time harmonic fields}\label{app:bispinor}
The bispinor formalism has much in common with wave-functions in quantum mechanics. In this supplementary, we will try to motivate the introduction of bispinors and show how they can be useful for describing the electromagnetic field. We will restrict our attention to time-harmonic (monochromatic) fields, which are fields with only a single frequency $\omega$. Any real monochromatic field (which we will denote with calligraphic letters, such as $\vc{\mathcal{A}}(t,\vc r)$) can be represented using a time-independent complex phasor $\vc A(\vc r)$ as:
\begin{equation}
    \vc{\mathcal{A}}(t,\vc r)=\Re[\vc A(\vc r)\ee^{-\ii\omega t}]
\end{equation}
Let us now consider some region in space $\mathcal{V}\subseteq\mathbb{R}^3$ that is occupied by a monochromatic electromagnetic field. In this region, we can calculate the total time-averaged energy of the field, which will be:
\begin{equation}
    \expval{E}=\frac{1}{4T}\int_0^T\qty[\iiint\nolimits_{\mathcal{V}}(\vc{\mathcal{E}}\vdot\vc{\mathcal{D}} +  \vc{\mathcal{H}}\vdot\vc{\mathcal{B}})\dd{V}]\dd{t} =\frac{1}{4}\iiint\nolimits_{\mathcal{V}} \Re(\vc{E}^*\!\vdot\vc{D} +  \vc{H}^*\!\vdot\vc{B})\dd{V}
\end{equation}
where $T=2\pi/\omega$ is the period of the wave, $\vc{\mathcal{E}}$ and $\vc{\mathcal{D}}$ are the electric field and displacement field (with phasor representations $\vc{{D}}$ and $\vc{{E}}$), and $\vc{\mathcal{H}}$ and $\vc{\mathcal{B}}$ are the magnetic field and flux density ($\vc{{B}}$ and $\vc{{H}}$ as phasors). For the purposes of this paper, we will assume a linear isotropic medium:
\begin{equation}
    \vc{D}=\varepsilon\vc{E}\,,\quad \vc{B}=\mu\vc{H}\,,
\end{equation}
however, the formalism can be generalised beyond this assumption. Physical fields will always have finite energy, even if we consider them across the whole space:
\begin{equation}\label{eq:finiteenergy}
    \expval{E}=\frac{1}{4}\iiint\nolimits_{\mathbb{R}^3} (\varepsilon\abs{\vc{E}(\vc{r})}^2 +  \mu\abs{\vc{H}(\vc{r})}^2)\dd{V}<\infty\,,
\end{equation}
This is not true for plane waves, but we should keep in mind that a physically realisable electromagnetic field will be approximately a plane wave only in a finite region of space. This situation is very similar to quantum mechanics where one has a wavefunction $\psi(\vc{r})$, which is a map that takes a position in space and returns a complex number $\mathbb{R}^3\to\mathbb{C}$. One can consider the energy eigenmode (solution to the time-independent Schr\"odinger equation $\hat{H}\psi(\vc{r})=E\psi(\vc{r})=\hbar\omega\psi(\vc{r})$), which is a direct analogue to the time-harmonic field, and show that the expectation value of the energy is as follows:
\begin{equation}\label{eq:qmexpect}
    \expval{E}=\iiint\nolimits_{\mathbb{R}^3} \psi^\dagger(\vc{r})\hat{H}\psi(\vc{r})\dd{V}=\hbar\omega\iiint\nolimits_{\mathbb{R}^3} \abs{\psi(\vc{r})}^2\dd{V}\,,
\end{equation}
where $\psi(\vc{r})\in L^2(\mathbb{R}^3,\mathbb{C})$ is a so called complex valued square integrable function, for which the integral converges
\begin{equation}
    \iiint\nolimits_{\mathbb{R}^3} \abs{\psi(\vc{r})}^2\dd{V}<\infty
\end{equation}
and therefore can be set to be equal to one. We say that $\psi(\vc{r})$ is normalisable and call $L^2(\mathbb{R}^3,\mathbb{C})$ the Hilbert space. Even in quantum mechanics, one often expresses solutions to Schr\"odinger equation in terms of functions that are formally not in the Hilbert space, keeping in mind that physical solutions have to be normalisable.

Knowing all this, we can ask a simple question: can we find a wavefunction $\vc\psi(\vc{r})$ such that \cref{eq:finiteenergy} can be written as \cref{eq:qmexpect}? Of course, that is possible, except this time we have $\vc{E}(\vc{r})$ and $\vc{H}(\vc{r})$, which are square integrable vector valued functions $L^2(\mathbb{R}^3,\mathbb{C}^3)$; that is, maps taking position and giving three-dimensional complex vector. That means that if we want a single ``wavefunction'', it has to be $\vc\psi(\vc{r})\in L^2(\mathbb{R}^3,\mathbb{C}^3)\oplus L^2(\mathbb{R}^3,\mathbb{C}^3)=L^2(\mathbb{R}^3,\mathbb{C}^6)$, so a square integrable map from position to \textit{two} three-dimensional complex vectors or one six-dimensional complex vectors. The $L^2(\mathbb{R}^3,\mathbb{C}^6)$ will play the role of our bispinor Hilbert space.

\subsection{Conventions}
The bispinor formalism has much in common with wave-functions in quantum mechanics (QM). As such one can think of it as being an abstract vector ($\ket{\psi}$ in Dirac notation) in Hilbert space which can be either expressed in position representation $\vc\psi(\vc{r})=\braket{\vc{r}}{\vc\psi}$ or in the momentum representation $\Tilde{\vc\psi}(\vc{p})=\braket{\vc{p}}{\vc\psi}$ (note that here $\vc{p}=\hbar\vc{k}$) which can be understood as two possible choices of basis in the subspace $L(\mathbb{R}^3)$. They are related by a Fourier transform:
\begin{equation}
    \vc\psi(\vc{r})=\iiint_{\mathbb{R}^3}\tilde{\vc\psi}(\vc{k})\,\ee^{\ii\vc{k}\vdot\vc{r}}\dd[3]{\vc{k}}
\end{equation}
\begin{equation}
    \ket{\psi}=\frac{A}{2\sqrt{\hbar\omega}}\iiint\limits_{\mathcal{V}} (\sqrt{\varepsilon} \vc{E}(\vc{r}) \otimes\uv{e}_\text{e} +  \sqrt{\mu} \vc{H}(\vc{r})\otimes\uv{e}_\text{m})\ket{\vc{r}}\dd{V}\,,
\end{equation}
this is the same as how one can in the conventional way either represent the fields by phasors $\vc{E}(\vc{r})$, $\vc{H}(\vc{r})$ or by their Fourier transforms $\tilde{\vc{E}}(\vc{k})$, $\tilde{\vc{H}}(\vc{k})$. In QM every observable is associated with a hermitian operator, for example the energy is associated with the Hamiltonian $\hat{H}$. The expectation value for energy in some region $\mathcal{V}$ is
\begin{equation}
    \expval{\hat{H}}{\psi}=\iiint\nolimits_{\mathcal{V}} \Re[\bar{\vc{\psi}}(\vc{r}) \hat{H}\vc{\psi}(\vc{r})]\dd{V}\,,
\end{equation}
where $\bar{\vc{\psi}}(\vc{r})=\braket{\vc\psi}{\vc{r}}$ is the vector dual to $\vc{\psi}(\vc{r})$ and we will discuss it further later. 
We know that for the electromagnetic field this can be calculated from the energy density as follows:
\begin{equation}\label{eq:hamiltonian_exp}
    \expval{\hat{H}}{\psi}=\iiint\nolimits_{\mathcal{V}} W(\vc{r})\dd{V}=\frac{1}{4}\iiint\nolimits_{\mathcal{V}} (\varepsilon \abs{\vc{E}}^2 + \mu \abs{\vc{H}}^2)\dd{V}\,,
\end{equation}
which means that we can identify the energy density with the following expression:
\begin{equation}
    W(\vc{r})=\Re[\bar{\vc{\psi}}(\vc{r}) \hat{H}\vc{\psi}(\vc{r})]\,.
\end{equation}
Same as in the case of QM we have some inherent ambiguity in defining the bispinor $\vc{\psi}(\vc{r})$. First, no observable or its expectation value should ever depend on the normalisation of this wave-function, so one can in principle choose it to be anything. However, the QM that most people are familiar with is such that the product $\braket{\psi}$ represents the probability of finding a particle in a given region $\mathcal{V}$ or alternatively $\Re[\bar{\vc{\psi}}(\vc{r})\vc{\psi}(\vc{r})]$ being the probability density. For a time-harmonic field the expected energy $\!\expval*{\hat{H}}=\hbar\omega N$, where $N$ is the number of photons in the region $\mathcal{V}$. This provides a very natural normalisation that is equivalent to the usual QM in the sense that all the operators in position representation have their usual form ($\hat{H}=\hbar\omega$, $\uv{p}=-\ii\hbar\grad$, \dots). The inner product is set to be the number of photons:
\begin{equation}
    \braket{\psi}=N=\frac{1}{4\hbar\omega}\iiint\nolimits_{\mathcal{V}} (\varepsilon \abs{\vc{E}}^2 + \mu \abs{\vc{H}}^2)\dd{V}=\iiint\nolimits_{\mathcal{V}} \frac{W(\vc{r})}{\hbar\omega}\dd{V}\,,
\end{equation}
where $W/\hbar\omega$ is the number density of photons. While this is a very natural choice in vacuum where it represents the number density of actual photons, it is not so in materials where this normalisation becomes somewhat arbitrary (since $W/\hbar\omega$ no longer represents number density of photons) and in both the main text and \cite{golat2024electromagnetic} we chose a different normalisation in which $\hat{H}=\hat{I}$ and $W(\vc{r})=\Re[\bar{\vc{\psi}}(\vc{r})\vc{\psi}(\vc{r})]$, in this supplementary however, we will be agnostic about this choice and instead use an arbitrary constant $A$ such that
\begin{equation}\label{eq:inner}
    \norm{\vc\psi}^2=\Re[\bar{\vc{\psi}}(\vc{r})\vc{\psi}(\vc{r})]=\tfrac{1}{4\hbar\omega}\abs{A}^2(\varepsilon \abs{\vc{E}}^2 + \mu \abs{\vc{H}}^2)=\tfrac{1}{\hbar\omega}\abs{A}^2W(\vc{r})\,,
\end{equation}
where $\abs{A}^2=1$ for the number of photons approach and $\abs{A}^2= \hbar\omega$ for the approach in the main text, and the operators will of course depend on this choice ($\hat{H}=\hbar\omega\abs{A}^{-2}$, $\uv{p}=-\ii\hbar\abs{A}^{-2}\grad$, \dots). Another ambiguity in representation comes from the choice of basis in the $\mathbb{C}^2$ space. In the main text, we choose an orthonormal basis $\lbrace\uv{e}_{\text{e}},\uv{e}_{\text{m}}\rbrace$.
\begin{equation}\label{eq:si_bispinor}
    \vc{\psi}(\vc{r})=\tfrac{A}{2\sqrt{\hbar\omega}} (\sqrt{\varepsilon} \vc{E}(\vc{r}) \otimes\uv{e}_\text{e} +  \sqrt{\mu} \vc{H}(\vc{r})\otimes\uv{e}_\text{m})\,,
\end{equation}
which has the advantage that the dual vector (conjugate under the inner product given by \cref{eq:inner}) is simply a complex conjugate with a transpose $\bar{\vc{\psi}}(\vc{r})=\vc{\psi}^\dagger(\vc{r})$. As we discussed in \cite{golat2024electromagnetic} probably a better choice for more complicated cases (such as when complex,  anisotropic or multiple media are involved) is the vacuum basis $\lbrace\uv{e}^0_{\text{e}},\uv{e}^0_{\text{m}}\rbrace$
\begin{equation}\label{eq:si_bispinor_vacuum}
    \vc{\psi}(\vc{r})=\tfrac{A}{2\sqrt{\hbar\omega}} (\sqrt{\varepsilon_0} \vc{E}(\vc{r}) \otimes\uv{e}^0_\text{e} +  \sqrt{\mu_0} \vc{H}(\vc{r})\otimes\uv{e}^0_\text{m})\,,
\end{equation}
in this case the dual vector is given by $\bar{\vc{\psi}}(\vc{r})=\vc{\gamma}^\dagger(\vc{r})$, where $\vc{\gamma}$ contains the auxiliary fields
\begin{equation}\label{eq:si_dual_bispinor_vacuum}
    \vc{\gamma}(\vc{r})=\tfrac{A}{2\sqrt{\hbar\omega}} ( \vc{D}(\vc{r})/ \sqrt{\varepsilon_0}\otimes\uv{e}^0_\text{e} +   \vc{B}(\vc{r})/\sqrt{\mu_0}\otimes\uv{e}^0_\text{m})\,.
\end{equation}
More details about different choices of bases in $\mathbb{C}^2$ can be found in the supplementary of \cite{golat2024electromagnetic}.
\subsection{Observables and projections}\label{app:operators}
Classically, observables associated with a time-harmonic field are usually understood as a time-averaged density. For example: energy density $W(\vc{r})$, linear momentum density $\vc{p}(\vc{r})$, orbital angular momentum density $\vc{L}(\vc{r})$, spin angular momentum density $\vc{S}(\vc{r})$ and many others. The way these can be understood in the bispinor formalism is as hermitian operators $\hat{O}$ (i.e. $\hat{W}$, $\uv{p}$,  $\uv{L}$, $\uv{S}$, \dots) for which the expectation value evaluated over some spatial region $\mathcal{V}$ is \begin{equation}
    \expval*{\hat{O}}=\expval{\hat{O}}{\psi}=\iiint\nolimits_{\mathcal{V}} \Re[\bar{\vc{\psi}}(\vc{r}) \hat{O}\vc{\psi}(\vc{r})]\dd{V}=\iiint\nolimits_{\mathcal{V}} O(\vc{r})\dd{V}\,,
\end{equation}
where $O(\vc{r})=\Re[\bar{\vc{\psi}}(\vc{r}) \hat{O}\vc{\psi}(\vc{r})]$ are precisely time-averaged densities of the observables in question. Since the bispinor itself is a vector in multiple vector spaces $\vc{\psi}(\vc{r})\in\mathbb{C}^2\times\mathbb{C}^3\times L^2(\mathbb{R}^3)$ the operators could in practice act on each of these vector spaces separately. The first example of a multiplicative operator we already saw in \cref{eq:hamiltonian_exp} which is the example of energy operator $\hat{W}=\hat{H}=\hbar\omega\abs{A}^{-2}$ which can in any representation be confirmed to lead to
\begin{equation}
 W=\Re(\bar{\vc{\psi}} \hat{W}\vc{\psi})=\frac{\Re[\bar{\vc{\psi}} (\hbar\omega)\vc{\psi}]}{\abs{A}^{2}}=\frac{1}{4}(\varepsilon \abs{\vc{E}}^2 + \mu \abs{\vc{H}}^2)\,.
\end{equation}
The next operator to consider is the canonical momentum operator, which only acts on the $\mathbb{C}^3\times L^2(\mathbb{R}^3)$ subspace as
\begin{equation}
 \vc{p}=\Re(\bar{\vc{\psi}} \uv{p}\vc{\psi})=\frac{\Re[\bar{\vc{\psi}} \vdot(-\ii\hbar\grad)\vc{\psi}]}{\abs{A}^{2}}=\frac{1}{4\omega}\Im[\varepsilon \vc{E}^*\vdot(\grad)\vc{E} + \mu \vc{H}^*\vdot(\grad)\vc{H}]\,.
\end{equation}
The orbital angular momentum operator is simply $\uv{L}=\vc{r}\cp\uv{p}$. Way more interesting is the spin angular momentum operator, which acts only on the subspace $\mathbb{C}^3$ and each of its components can be written as 
\begin{equation}
    \hat{S}_x=\frac{\ii\hbar\uv{x}\cp}{\abs{A}^{2}}\,,\quad\hat{S}_y=\frac{\ii\hbar\uv{y}\cp}{\abs{A}^{2}}\,,\quad\hat{S}_z=\frac{\ii\hbar\uv{z}\cp}{\abs{A}^{2}}\,,
\end{equation}
and using the triple product identity $\vc{A}\vdot(\vc{B}\cp\vc{C})=-\vc{B}\vdot(\vc{A}\cp\vc{C})$ we can confirm that 
\begin{equation}
 \vc{S}=\Re(\bar{\vc{\psi}} \uv{S}\vc{\psi})=\frac{\Re[\bar{\vc{\psi}} (-\ii\hbar\cp)\vc{\psi}]}{\abs{A}^{2}}=\frac{1}{4\omega}\Im(\varepsilon \vc{E}^*\cp\vc{E} + \mu \vc{H}^*\cp\vc{H})\,.
\end{equation}
Notice that neither of the previous operators affected the $\mathbb{C}^2$ subspace yet. As a consequence all previous observables are always a sum of separate electric and magnetic contributions. In contrast, operators that act on $\mathbb{C}^2$ can produce mixed terms that depend on both the electric and magnetic field. To understand this better let's start with projection operators. As discussed in the main text and previous section one can find three pairs orthonormal basis vectors $\lbrace\uv{e}_{\text{e}},\uv{e}_{\text{m}}\rbrace$ and
\begin{equation}\label{eq:bases_si}
    \uv{e}_\text{p/a}=\tfrac{1}{\sqrt{2}}(\uv{e}_\text{e}\pm\uv{e}_\text{m}),\quad
    \uv{e}_\text{R/L}=\tfrac{1}{\sqrt{2}}(\uv{e}_\text{e}\mp\ii\uv{e}_\text{m})\,.
\end{equation}
In either of these bases one can write the bispinor as
\begin{equation}
    \vc{\psi}(\vc{r})= \vc{F}_i(\vc{r}) \otimes\uv{e}_i +  \vc{F}_j(\vc{r})\otimes\uv{e}_j\,,
\end{equation}
These basis vectors allow for definition of projection operators $\hat{P}_i=\uv{e}^{\phantom{\dagger}}_i\uv{e}_i^\dagger$ where $i\in\{\text{e,m,p,a,}\textsc{r,l}\}$ with action
\begin{equation}
    \hat{P}_i\vc{\psi}=(\uv{e}_i^*\vdot\vc{\psi})\otimes\uv{e}^{\phantom{\dagger}}_i=\vc{F}_i(\vc{r}) \otimes\uv{e}_i\,.
\end{equation}
These projection operators in turn can be used to define a basis $\lbrace\hat{P}_0, \hat{P}_1, \hat{P}_2, \hat{P}_3\rbrace$ for any $\mathbb{C}^2$ operator.
We can start by defining the $\hat{P}_0$ operator as:
\begin{equation}
    \hat{P}_0=\hat{P}_\text{e}+\hat{P}_\text{m}=\hat{P}_\text{a}+\hat{P}_\text{p}=\hat{P}_\text{R}+\hat{P}_\text{L}
\end{equation}
which due to the completeness of bases $\lbrace\uv{e}_{\text{e}},\uv{e}_{\text{m}}\rbrace$, $\lbrace\uv{e}_{\text{p}},\uv{e}_{\text{a}}\rbrace$ and $\lbrace\uv{e}_{\text{R}},\uv{e}_{\text{L}}\rbrace$
acts as the identity operator $\hat{P}_0\vc{\psi}=\vc{\psi}$. The remaining are
\begin{equation}
        \hat{P}_1=\hat{P}_\text{e}-\hat{P}_\text{m}\,,\quad\quad
        \hat{P}_2=\hat{P}_\text{a}-\hat{P}_\text{p}\,,\quad\quad
        \hat{P}_3=\hat{P}_\text{R}-\hat{P}_\text{L}\,.
\end{equation}
By explicitly choosing a representation for the $\mathbb{C}^2$ space, or by using \cref{eq:bases_si} to prove that $\comm*{\hat{P}_A}{ \hat{P}_B}=2 \ii \epsilon_{ABC} \hat{P}_C$ for $A,B,C\in\{1,2,3\}$ one can show that these operators are generators of of Lie Algebra $\mathfrak{su}(2)$ or in other words they can be represented using Pauli matrices, see \cref{tab:transposed_matrix_basis}. 

By combining these operators together with the more conventional observables 
(like $\hat{W}$, $\uv{p}$,  $\uv{L}$, $\uv{S}$, \dots) one can obtain new observables. Consider an observable $\hat{O}$ combined with $\hat{P}_\text{e}$
\begin{equation}
    O_\text{e}=\Re(\bar{\vc{\psi}} \hat{O}\hat{P}_\text{e}\vc{\psi})=\Re[\bar{\vc{\psi}} \hat{O}\vdot(\uv{e}_\text{e}\otimes\uv{e}_\text{e}^\ast)\vdot\vc{\psi}]=\Re[(\bar{\vc{\psi}} \vdot\uv{e}_\text{e})\hat{O}(\uv{e}_\text{e}\mathclap{^{\!\!*}}\vdot\vc{\psi})]=\tfrac{1}{4\hbar\omega}\abs{A}^2\Re(\varepsilon \vc{E}^*\hat{O}\vc{E})\,,
\end{equation}
here we assumed that $\hat{O}$ itself does not act on the $\mathbb{C}^2$ subspace and hence commutes with $\uv{e}_\text{e}$. The same is true for all the other projection operators. That means that the energy, momentum, orbital and spin angular momentum of each projection $i\in\{\text{e,m,p,a,}\textsc{r,l}\}$ can be written as
\begin{equation}
    \begin{split}
        W_i&=\Re(\bar{\vc{\psi}} \hat{W}\hat{P}_i\vc{\psi})=\hbar\omega\norm{\uv{e}_i\mathclap{^{\!\!*}}\vdot\vc\psi}^2/\abs{A}^{2}\,,\\
        \vc{p}_i&=\Re(\bar{\vc{\psi}} \uv{p}\hat{P}_i\vc{\psi})={\Re[(\bar{\vc{\psi}}\vdot\uv{e}_i) (-\ii\hbar\grad)(\uv{e}_i\mathclap{^{\!\!*}}\vdot\vc{\psi})]}/\abs{A}^{2}\,,\\
        \vc{L}_i&=\vc{r}\cp\Re(\bar{\vc{\psi}} \uv{p}\hat{P}_i\vc{\psi})=\vc{r}\cp{\Re[(\bar{\vc{\psi}}\vdot\uv{e}_i) (-\ii\hbar\grad)(\uv{e}_i\mathclap{^{\!\!*}}\vdot\vc{\psi})]}/\abs{A}^{2}\,,\\
        \vc{S}_i&=\Re(\bar{\vc{\psi}} \uv{S}\hat{P}_i\vc{\psi})={\Re[(\bar{\vc{\psi}}\vdot\uv{e}_i) (-\ii\hbar\cp)(\uv{e}_i\mathclap{^{\!\!*}}\vdot\vc{\psi})]}/\abs{A}^{2}\,,
        \end{split}
\end{equation}
By combining the basis of operators on $\mathbb{C}^2$ (i.e. $P_A$ with $A\in\{0,1,2,3\}$) with operators $\hat{O}$ representing observables (that is $\hat{W}$, $\uv{p}$,  $\uv{L}$ and $\uv{S}$) we get four distinct versions $\{\hat{O}_0, \hat{O}_1, \hat{O}_2, \hat{O}_3\}$:
\begin{equation}
    \begin{alignedat}{2}
        W_A&=\Re(\bar{\vc{\psi}} \hat{W}\hat{P}_A\vc{\psi})=\frac{\Re[\bar{\vc{\psi}}(\hbar\omega)\hat{P}_A\vc{\psi}]}{\abs{A}^{2}}\,,\\
        \vc{p}_A&=\Re(\bar{\vc{\psi}} \uv{p}\hat{P}_A\vc{\psi})=\frac{\Re[\bar{\vc{\psi}}\vdot(-\ii\hbar\grad)\hat{P}_A\vc{\psi}]}{\abs{A}^{2}}\,,\\
        \vc{L}_A&=\vc{r}\cp\Re(\bar{\vc{\psi}} \uv{p}\hat{P}_A\vc{\psi})=\frac{\Re[\bar{\vc{\psi}}\vdot(-\ii\hbar\vc{r}\cp\grad)\hat{P}_A\vc{\psi}]}{\abs{A}^{2}}\,,\\
        \vc{S}_A&=\Re(\bar{\vc{\psi}} \uv{S}\hat{P}_A\vc{\psi})=\frac{\Re[\bar{\vc{\psi}}(-\ii\hbar\cp)\hat{P}_A\vc{\psi}]}{\abs{A}^{2}}\,,
        \end{alignedat}
\end{equation}
which we have introduced in \cite{golat2024electromagnetic} and in the main text, and which represent sums and differences of these observables for different projections. 

\section{Dipole interactions in bispinor formalism}\label{app:operatorPHFG}
\subsection{Basis independent expressions for power, helicity, force and torque}
In a similar manner to the fields, one can define a bispinor that contains the dipoles
\begin{equation}\label{eq:si_bispinor_dipole}
    \vc{\pi}(\vc{r})=\tfrac{A}{2\sqrt{\hbar\omega}} ( \vc{p}(\vc{r})/\sqrt{\varepsilon} \otimes\uv{e}_\text{e} +  \sqrt{\mu} \vc{m}(\vc{r})\otimes\uv{e}_\text{m})\,,
\end{equation}
note that same as in the case of the fields one does not have to use this particular orthonormal basis. Having defined this bispinor we can rewrite the dipolar absorbed power
\begin{equation}
   \begin{split}
       \mathcal{P}_\text{abs}&=\frac{\omega}{2}\Big[\Im\qty(\vc{E}^\ast\!\vdot\vc{p}+\mu\vc{H}^\ast\!\vdot\vc{m})-\frac{k^3}{6\pi}\Big(\frac{1}{\varepsilon}\abs{\vc{p}}^2+\mu\abs{\vc{m}}^2\Big)\Big]\\
       &=\frac{\omega}{2}\Big[\Im\Big(\sqrt{\varepsilon}\vc{E}^\ast\!\vdot\frac{\vc{p}}{\sqrt{\varepsilon}}+\sqrt{\mu}\vc{H}^\ast\!\vdot\sqrt{\mu}\vc{m}\Big)-\frac{k^3}{6\pi}\Big(\frac{\abs{\vc{p}}^2}{\varepsilon}+\mu\abs{\vc{m}}^2\Big)\Big]\\
       &=\frac{\omega}{2}\frac{4\hbar\omega}{\abs{A}^2}\Big[\Im(\bar{\vc{\psi}}\vdot\vc{\pi})-\frac{k^3}{6\pi}(\bar{\vc{\pi}}\vdot\vc{\pi})\Big]=\frac{2kc}{\abs{A}^2}\Big[\Im(\bar{\vc{\psi}}(\hbar\omega)\vc{\pi})-\frac{k^3}{6\pi}(\bar{\vc{\pi}}(\hbar\omega)\vc{\pi})\Big]\\
       &=2k\,\Im(\bar{\vc{\psi}} c\hat{W}_0\vc{\pi})-\frac{k^4}{3\pi}(\bar{\vc{\pi}} c\hat{W}_0\vc{\pi})\,.
   \end{split}
\end{equation}
This expression is independent of the choice of basis and normalisation. Next we can rewrite the absorbed helicity
\begin{equation}
    \begin{split}
    \dot{\mathcal{H}}_\text{abs}&=\frac{\sqrt{\varepsilon\mu}}{2}\Big[\Re\Big(\vc{E}^\ast\!\vdot\vc{m}-\frac{1}{\varepsilon}\vc{H}^\ast\!\vdot\vc{p}\Big)-\frac{k^3}{3\pi}\Im\Big(\frac{1}{\varepsilon}\vc{p}\vdot\vc{m}^\ast\Big)\Big]\\
    &=\frac{1}{2}\Big[\Im\Big(\sqrt{\varepsilon}\vc{E}^\ast\!\vdot\ii\sqrt{\mu}\vc{m}-\sqrt{\mu}\vc{H}^\ast\!\vdot\ii\frac{\vc{p}}{\sqrt{\varepsilon}}\Big)-\frac{k^3}{6\pi}\Big(\frac{\vc{p}^\ast}{\sqrt{\varepsilon}}\vdot\ii\sqrt{\mu}\vc{m}-\sqrt{\mu}\vc{m}^\ast\vdot\ii\frac{\vc{p}}{\sqrt{\varepsilon}}\Big)\Big]\\
    &=\frac{2\hbar\omega}{\abs{A}^2}\Big[\Im(\bar{\vc{\psi}}\hat{P}_3\vc{\pi})-\frac{k^3}{6\pi}(\bar{\vc{\pi}}\hat{P}_3\vc{\pi})\Big]=2\,\Im(\bar{\vc{\psi}} \hat{W}_3\vc{\pi})-\frac{k^3}{3\pi}(\bar{\vc{\pi}} \hat{W}_3\vc{\pi})\,.
    \end{split}
\end{equation}
In the third line we used the fact that $\hat{P}_3$ is represented by Pauli matrix $-\tens{\sigma}_2$ in electromagnetic basis. The force is
\begin{equation}
\begin{split}
    {\vc{F}}&=\frac{1}{2}\Re\Big[\vc{p}^\ast\!\vdot(\grad)\vc{E}+\mu\vc{m}^\ast\!\vdot(\grad)\vc{H}-\frac{k^4}{6\pi}\sqrt{\frac{\mu}{\varepsilon}}(\vc{p}\cp\vc{m}^\ast)\Big]\\
    &=\frac{1}{2}\Im\Big[\frac{\vc{p}^\ast}{\sqrt{\varepsilon}}\vdot(-\ii\grad)\sqrt{\varepsilon}\vc{E}+\sqrt{\mu}\vc{m}^\ast\!\vdot(-\ii\grad)\sqrt{\mu}\vc{H}\Big]-\frac{k^4}{24\pi}(-\ii)\Big(\frac{\vc{p}^\ast}{\sqrt{\varepsilon}}\cp\ii\sqrt{\mu}\vc{m}-\sqrt{\mu}\vc{m}^\ast\cp\ii\frac{\vc{p}}{\sqrt{\varepsilon}}\Big)\\
    &=\frac{2\hbar\omega}{\abs{A}^2}\Big[\Im(\bar{\vc{\pi}}\vdot(-\ii\grad)\vc{\psi})-\frac{k^4}{12\pi}(\bar{\vc{\pi}}(-\ii\cp)\hat{P}_3\vc{\pi})\Big]=\frac{2kc}{\abs{A}^2}\Big[\Im(\bar{\vc{\pi}}\vdot(-\ii\hbar\grad)\vc{\psi})-\frac{k^4}{12\pi}(\bar{\vc{\pi}}(-\ii\hbar\cp)\hat{P}_3\vc{\pi})\Big]
    \\
    &=2k\,\Im(\bar{\vc{\pi}} c\uv{p}_0\,\vc{\psi})-\frac{k^5}{6\pi }(\bar{\vc{\pi}} c\uv{S}_3\vc{\pi})\,.
\end{split}
\end{equation}
Finally, the torque will be
\begin{equation*}
\begin{split}
    {\vc{\varGamma}}&=\frac{1}{2}\Re\Big(\vc{p}^\ast\!\cp\vc{E}+\!\mu\vc{m}^\ast\!\cp\vc{H}\Big)
    -\frac{k^3}{12\pi}\Im\Big(\frac{1}{\varepsilon}\vc{p}^\ast\!\!\cp\!\vc{p}+\mu\vc{m}^\ast\!\!\cp\!\vc{m}\Big)\\
    &=\frac{1}{2}\Im\Big(\frac{\vc{p}^\ast}{\sqrt{\varepsilon}}(-\ii\cp)\sqrt{\varepsilon}\vc{E}+\sqrt{\mu}\vc{m}^\ast(-\ii\cp)\sqrt{\mu}\vc{H}\Big)
    -\frac{k^3}{12\pi}\Big(\frac{\vc{p}^\ast}{\sqrt{\varepsilon}}(-\ii\cp)\frac{\vc{p}}{\sqrt{\varepsilon}}+\sqrt{\mu}\vc{m}^\ast(-\ii\cp)\sqrt{\mu}\vc{m}\Big)\\
    &=\frac{2\hbar\omega}{\abs{A}^2}\Big[\Im(\bar{\vc{\pi}}(-\ii\cp)\vc{\psi})-\frac{k^3}{6\pi}(\bar{\vc{\pi}}(-\ii\cp)\vc{\pi})\Big]=\frac{2kc}{\abs{A}^2}\Big[\Im(\bar{\vc{\pi}}(-\ii\hbar\cp)\vc{\psi})-\frac{k^3}{6\pi}(\bar{\vc{\pi}}(-\ii\hbar\cp)\vc{\pi})\Big]\\
    &=2k\,\Im(\bar{\vc{\pi}} c\uv{S}_0\,\vc{\psi})-\tfrac{k^4}{3\pi }(\bar{\vc{\pi}} c\uv{S}_0\vc{\pi})\,.
\end{split}
\end{equation*}
\section{Cross sections for bi-isotropic dipole}
\subsection{Power of linear bi-isotropic dipole}\label{app:power}
In this section, we shall derive extinction and scattering cross sections for a linear bi-isotropic dipole $\vc{\pi}=\tens{A}\vc{\psi}$. We start by writing the extinction power in the basis independent way:
\begin{equation}\label{eq:powex}
   \begin{split}
       \mathcal{P}_\text{ext}&=2k\,\Im(\bar{\vc{\psi}} c\hat{W}_0\vc{\pi})=2\omega\,\Im(\bar{\vc{\psi}} \hat{W}_0\tens{A}\vc{\psi})\,,
   \end{split}
\end{equation}
then we decompose the polarisability operator into the basis of projection operators (Pauli matrices)
\begin{equation} \label{eq:Adec}
    \tens{A}=\tfrac{1}{2}(\alpha_0\hat{P}_0+\alpha_1\hat{P}_1+\alpha_2\hat{P}_2+\alpha_3\hat{P}_3)\,.
\end{equation}
Remember that $\hat{W}_0\hat{P}_A=\hat{W}_A$ and $W_A=\bar{\vc{\psi}} \hat{W}_A\vc{\psi}$, plugging \cref{eq:Adec} back to \cref{eq:powex} we get:
\begin{equation}
   \begin{split}
       \mathcal{P}_\text{ext}&=\sum_{A=0}^3\,\omega\Im(\bar{\vc{\psi}} \alpha_A\hat{W}_A\vc{\psi})=\sum_{A=0}^3\,\omega\Im(\alpha_A\bar{\vc{\psi}} \hat{W}_A\vc{\psi})=\sum_{A=0}^3\,\omega\Im(\alpha_A){W}_A=\sum_{A=0}^3\,\sigma^A_\text{ext}c{W}_A\,.,
   \end{split}
\end{equation}
where the last equal sign can be understood as a definition of extinction power cross sections $\sigma^A_\text{ext}$.
The power scattered by the particle is quadratic in polarisabilities:
\begin{equation}
   \begin{split}
       \mathcal{P}_\text{sca}&=\frac{k^4}{3\pi}(\bar{\vc{\pi}} c\hat{W}_0\vc{\pi})=\frac{c k^4}{3\pi}(\bar{\vc{\psi}} \tens{A}^\dagger\hat{W}_0\tens{A}\vc{\psi})=\frac{c k^4}{3\pi}(\bar{\vc{\psi}} \tens{A}^\dagger\tens{A}\hat{W}_0\vc{\psi})\,,
   \end{split}
\end{equation}
so to simplify this derivation let us use a little trick and define a vector-valued operator $\uv{P}=(\hat{P}_1,\hat{P}_2,\hat{P}_3)^\intercal$, real energy asymmetry vector $\vc{W}=(W_1,W_2,W_3)^\intercal$ and complex polarisation asymmetry vector $\vc{\alpha}=(\alpha_1,\alpha_2,\alpha_3)^\intercal$. The polarisability decomposition in the basis is then 
\begin{equation}
    \tens{A}=\tfrac{1}{2}(\alpha_0\hat{P}_0+\vc{\alpha}\vdot\uv{P})\,,
\end{equation}
which we can use to write the product (we used $\hat{P}_0^2=\hat{P}_0$ and $\hat{P}_A^\dagger=\hat{P}_A$)
\begin{equation}
    \tens{A}^\dagger\tens{A}=\tfrac{1}{4}[|\alpha_0|^2\hat{P}_0+(\alpha_0^*\vc{\alpha}+\alpha_0\vc{\alpha}^*)\vdot\uv{P}+
    (\vc{\alpha}^*\vdot\uv{P})\vdot(\vc{\alpha}\vdot\uv{P})]\,,
\end{equation}
remembering that $\hat{P}_A$ are just Pauli matrices, we can use the straightforward to check identity:
\begin{equation}
    (\vc{a}\vdot\uv{P})(\vc{b}\vdot\uv{P})=(\vc{a}\vdot\vc{b})\hat{P}_0+\ii(\vc{a}\cp\vc{b})\vdot\uv{P}\,,
\end{equation}
we can also use the fact that $\ii
    \vc{\alpha}^*\cp\vc{\alpha}=-\Im(
    \vc{\alpha}^*\cp\vc{\alpha})$, which leads to the quadratic operator being:
\begin{equation}
    \tens{A}^\dagger\tens{A}=\tfrac{1}{4}[(|\alpha_0|^2+|\vc{\alpha}|^2)\hat{P}_0+[2\Re(\alpha_0^*\vc{\alpha})-\Im(
    \vc{\alpha}^*\cp\vc{\alpha})]\vdot\uv{P}]\,
\end{equation}
Finally, we can write the scattered power as:
\begin{equation}\label{eq:scatcs}
   \begin{split}
       \mathcal{P}_\text{sca}&=\frac{k^4}{12\pi}\qty{(|\alpha_0|^2+|\vc{\alpha}|^2)cW_0+[2\Re(\alpha_0^*\vc{\alpha})-\Im(
    \vc{\alpha}^*\cp\vc{\alpha})]\vdot c\vc{W}}=\sum_{A=0}^3\,\sigma^A_\text{sca}c{W}_A\,.
   \end{split}
\end{equation}
The absorbed power can be obtained from the difference $\mathcal{P}_\text{abs}=\mathcal{P}_\text{ext}-\mathcal{P}_\text{sca}$ or in the operator form 
\begin{equation}
    \mathcal{P}_\text{abs}=2\omega\qty[\bar{\vc{\psi}}\qty( \frac{\tens{A}-\tens{A}^\dagger}{2\ii}-\frac{k^3}{6\pi}\tens{A}^\dagger\tens{A})\hat{W}_0\vc{\psi}]\,,
\end{equation}
which for a passive particle has to obey $\mathcal{P}_\text{abs}\geq0$, this means that the operator 
\begin{equation}
       \tens{\sigma}_\text{abs}=2k\qty( \frac{\tens{A}-\tens{A}^\dagger}{2\ii}-\frac{k^3}{6\pi}\tens{A}^\dagger\tens{A})\succeq0\,,
\end{equation}
has to be positive semidefinite. That happens if all eigenvalues $\lambda_i(\tens{\sigma}_\text{abs})\geq0$. We can decompose this operator into the Pauli matrices again
\begin{equation}
       \tens{\sigma}_\text{abs}={\sigma}_\text{abs}^0\hat{P}_0+\vc{{\sigma}}_\text{abs}\vdot\uv{P}\,,
\end{equation}
where ${\sigma}_\text{abs}^0$ is the total absorption cross section and $\vc{{\sigma}}_\text{abs}=({{\sigma}}_\text{abs}^1,{{\sigma}}_\text{abs}^2,{{\sigma}}_\text{abs}^3)^\intercal$ is its asymmetry:
\begin{equation}
\begin{split}
    {\sigma}_\text{abs}^0&=k\Im\alpha_0-\frac{k^4}{12\pi}(|\alpha_0|^2+|\vc{\alpha}|^2)\\
    \vc{{\sigma}}_\text{abs}&=k\Im\vc{\alpha}-\frac{k^4}{12\pi}[2\Re(\alpha_0^*\vc{\alpha})-\Im(
    \vc{\alpha}^*\cp\vc{\alpha})]
\end{split}
\end{equation}
since this is a 2-by-2 matrix, the eigenvalues are of the form $\lambda_\pm={\sigma}_\text{abs}^0\pm|\vc{{\sigma}}_\text{abs}|\geq0$ both of which have to be non-negative, but since one of the two conditions is weaker than the other, we end up with the condition:
\begin{equation}
    {\sigma}_\text{abs}^0\geq|\vc{{\sigma}}_\text{abs}|\,,
\end{equation}
which means that the absorption cross sections of a passive particle form a symmetry sphere (Bloch sphere) of radius ${\sigma}_\text{abs}^0$, such that the vector $\vc{{\sigma}}_\text{abs}$ has to be inside this sphere. A particle with $\vc{{\sigma}}_\text{abs}$ located exactly at the surface will be perfectly asymmetric, and there will be some illumination with $\vc{W}=(W_1,W_2,W_3)^\intercal$ and $|\vc{W}|=W_0$ antiparallel to $\vc{{\sigma}}_\text{abs}$ that gives exactly $\mathcal{P}_\text{abs}=0$.

\begin{equation}
       \mathcal{P}_\text{abs}=c({\sigma}_\text{abs}^0W_0+\vc{{\sigma}}_\text{abs}\vdot\vc{W})\,,
\end{equation}

% $$\Im{\alpha_0}-g(|\alpha_0|^2+|\vc{\alpha}|^2)\geq |\Im{\vc{\alpha}}-g[2\Re(\alpha_0^*\vc{\alpha})-\Im(
%     \vc{\alpha}^*\cp\vc{\alpha})]|$$

\subsection{Power cross sections}
These cross sections can be calculated in terms of complex polarisabilities. Let us start with the extinction power cross section, it can be written very simply as $\sigma_\text{ext}^A=k\Im{\alpha_A}$, where $A\in\{0,1,2,3\}$ or $A\in\{\text{e,m,p,a},\textsc{r,l}\}$ or explicitly:
\begin{alignat}{3}
% \begin{equation}
%     \begin{alignedat}{2}
        \sigma_\text{ext}^0=k\Im\alpha_0&=k\Im(\alpha_\text{e}+\alpha_\text{m})&&=\tfrac{1}{2}(\sigma_\text{ext}^\text{e}+\sigma_\text{ext}^\text{m})\notag\\&=k\Im(\alpha_\text{a}+\alpha_\text{p})&&=\tfrac{1}{2}(\sigma_\text{ext}^\text{a}+\sigma_\text{ext}^\text{p})\\&=k\Im(\alpha_\text{R}+\alpha_\text{L})&&=\tfrac{1}{2}(\sigma_\text{ext}^\text{R}+\sigma_\text{ext}^\text{L})\,,\notag\\
%     \end{alignedat}
% \end{equation}
% the remaining extinction power cross sections will represent asymmetry in the extinction
% \begin{alignat}{3}
    \sigma_\text{ext}^1=k\Im\alpha_1&=k\Im(\alpha_\text{e}-\alpha_\text{m})&&=\tfrac{1}{2}(\sigma_\text{ext}^\text{e}-\sigma_\text{ext}^\text{m})\,,\\
    \sigma_\text{ext}^2=k\Im\alpha_2&=k\Im(\alpha_\text{a}-\alpha_\text{p})&&=\tfrac{1}{2}(\sigma_\text{ext}^\text{a}-\sigma_\text{ext}^\text{p})\,,\\
    \sigma_\text{ext}^3=k\Im\alpha_3&=k\Im(\alpha_\text{R}-\alpha_\text{L})&&=\tfrac{1}{2}(\sigma_\text{ext}^\text{R}-\sigma_\text{ext}^\text{L})\,.
\end{alignat}
The last equality comes from the fact that if we choose an illumination that maximally breaks a certain symmetry, for instance, left-handed illumination with $W_0=W_\text{L}$ and $W_3=-W_\text{L}$, then the extinction power will be
\begin{equation*}
    \mathcal{P}_\text{ext}=\sum_{A=0}^3\sigma_\text{ext}^AcW_A=(\sigma_\text{ext}^0-\sigma_\text{ext}^3)cW_\text{L}=\sigma_\text{ext}^\text{L}cW_\text{L}\,,
\end{equation*}
where $\sigma_\text{ext}^\text{L}$ is precisely the extinction cross section that would be measured using left-handed illumination.
The scattering power cross sections $\sigma_\text{sca}^A$ with $A\in\{0,1,2,3\}$ can be obtained from \cref{eq:scatcs} as:
\begin{align}
    \sigma_\text{sca}^0&=\tfrac{k^4}{12\pi}(\abs{\alpha_0}^2+\abs{\alpha_1}^2+\abs{\alpha_2}^2+\abs{\alpha_3}^2),\\
    \sigma_\text{sca}^1&=\tfrac{k^4}{6\pi}[\Re(\alpha_0^\ast\alpha_1)-\Im(\alpha_2^\ast\alpha_3)]\,,\\
    \sigma_\text{sca}^2&=\tfrac{k^4}{6\pi}[\Re(\alpha_0^\ast\alpha_2)-\Im(\alpha_3^\ast\alpha_1)]\,,\\
    \sigma_\text{sca}^3&=\tfrac{k^4}{6\pi}[\Re(\alpha_0^\ast\alpha_3)-\Im(\alpha_1^\ast\alpha_2)]\,,
\end{align}
and we can use the same logic as before to write cross section for each projection as: 
\begin{alignat}{3}
    \sigma_\text{sca}^\text{R/L}&=\sigma_\text{sca}^0\pm\sigma_\text{sca}^3&&=\tfrac{k^4}{3\pi}(\abs*{\alpha_\text{R/L}}^2&&+\tfrac14\abs{\alpha_1\pm\ii\alpha_2}^2),\\
    \sigma_\text{sca}^\text{e/m}&=\sigma_\text{sca}^0\pm\sigma_\text{sca}^1&&=\tfrac{k^4}{3\pi}(\abs*{\alpha_\text{e/m}}^2&&+\tfrac14\abs{\alpha_2\pm\ii\alpha_3}^2),\\
    \sigma_\text{sca}^\text{a/p}&=\sigma_\text{sca}^0\pm\sigma_\text{sca}^2&&=\tfrac{k^4}{3\pi}(\abs*{\alpha_\text{a/p}}^2&&+\tfrac14\abs{\alpha_3\pm\ii\alpha_1}^2).
\end{alignat}
Notice that while it is usually believed that $\sigma_\text{sca}^\text{e/m}\propto\abs*{\alpha_\text{e/m}}^2$ that is only true if the particle is achiral and reciprocal.

\subsection{Helicity cross sections}
The helicity extinction cross section 
and its asymmetry contributions can be written as follows
\begin{alignat}{3}
    \gamma_\text{ext}^0&=k\Im\alpha_3&&=k\Im(\alpha_\text{R}-\alpha_\text{L})&&=\tfrac{1}{2}(\gamma_\text{ext}^\text{R}+\gamma_\text{ext}^\text{L})\,,\\
    \mathop{-}\gamma_\text{ext}^1&=k\Re\alpha_2&&=k\Re(\alpha_\text{a}-\alpha_\text{p})\,,\\
    \gamma_\text{ext}^2&=k\Re\alpha_1&&=k\Re(\alpha_\text{e}-\alpha_\text{m})\,,\\
    \gamma_\text{ext}^3&=k\Im\alpha_0&&=k\Im(\alpha_\text{R}+\alpha_\text{L})&&=\tfrac{1}{2}(\gamma_\text{ext}^\text{R}-\gamma_\text{ext}^\text{L})\,.
\end{alignat}
Notice that $\gamma_\text{ext}^{3}=\sigma_\text{ext}^{0}$ and $\gamma_\text{ext}^{0}=\sigma_\text{ext}^{3}$, while the scattered helicity cross sections will be
\begin{align}
    \gamma_\text{sca}^0&=\tfrac{k^4}{6\pi}[\Re(\alpha_3^\ast\alpha_0)+\Im(\alpha_1^\ast\alpha_2)],\\
    \gamma_\text{sca}^1&=\tfrac{k^4}{6\pi}[\Re(\alpha_3^\ast\alpha_1)+\Im(\alpha_0^\ast\alpha_2)]\,,\\
    \gamma_\text{sca}^2&=\tfrac{k^4}{6\pi}[\Re(\alpha_3^\ast\alpha_2)+\Im(\alpha_1^\ast\alpha_0)]\,,\\
    \gamma_\text{sca}^3&=\tfrac{k^4}{12\pi}(\abs{\alpha_0}^2-\abs{\alpha_1}^2-\abs{\alpha_2}^2+\abs{\alpha_3}^2)\,,
\end{align}
and as in the case of extinction, we can write the right- and left- handed helicity scattering cross sections 
\begin{alignat}{2}
    \gamma_\text{sca}^\text{R/L}=\pm\tfrac{k^4}{3\pi}(\abs*{\alpha_\text{R/L}}^2-\tfrac14\abs{\alpha_1\pm\ii\alpha_2}^2).
\end{alignat}

\section{Chiral basis for electromagnetic radiation}\label{app:handedness}
In the main text, we assert that each handedness of light possesses its own distinct energy density and flux. This section provides a mathematical justification for that claim. We begin by expressing the electric field as a superposition of plane waves, i.e. its inverse Fourier transform, which is always possible. For every propagation direction $\uv{k}$ one may introduce local circular-polarisation unit vectors $\uv{e}_\text{R/L}(\uv{k})$. These are defined only up to a complex phase factor $\ee^{\mp\ii\phi}$, reflecting the freedom to rotate the local transverse frame by an arbitrary angle $\phi$ about $\uv{k}$. We can therefore write:
\begin{equation}\label{eq:ERL}
\begin{split}
    \vc{E}(\vc{r})&=\!\!\iiint\tilde{\vc{E}}(\vc{k})\ee^{\ii\vc{k}\vdot\vc{r}}\dd[3]k=\!\!\iiint[\tilde{\vc{E}}_\text{R}(\vc{k})\uv{e}_\text{R}(\uv{k})+\tilde{\vc{E}}_\text{L}(\vc{k})\uv{e}_\text{L}(\uv{k})]\ee^{\ii\vc{k}\vdot\vc{r}}\dd[3]k=\vc{E}_\text{R}(\vc{r})+\vc{E}_\text{L}(\vc{r})\,,
\end{split}
\end{equation}
and although the basis vectors $\uv{e}_{\text{R/L}}(\uv{k})$ are not unique, the corresponding fields $\vc{E}_{\text{R/L}}(\vc{r})$ are uniquely defined. Local phase ambiguity can always be absorbed into the spectral amplitudes $\tilde{\vc{E}}_\text{R/L}(\vc{k})$, leaving the physical fields themselves invariant. We can write $\vc{E}(\vc{r})=\vc{E}_\text{R}(\vc{r})+\vc{E}_\text{L}(\vc{r})$ thanks to the linearity of integration. We can find the magnetic fields using the Maxwell-Faraday equation $\curl{\vc{E}}=\ii\eta k\vc{H}$ leading to a similar split: 
\begin{equation}\label{eq:HRL}
\begin{split}
    \!\!\vc{H}(\vc{r})&=\frac{1}{\eta}\!\iiint\uv{k}\cp\tilde{\vc{E}}(\vc{k})\ee^{\ii\vc{k}\vdot\vc{r}}\dd[3]k=\frac{1}{\eta}\!\iiint[-\ii\tilde{\vc{E}}_\text{R}\uv{e}_\text{R}(\vc{k})+\ii\tilde{\vc{E}}_\text{L}\uv{e}_\text{L}(\vc{k})]\ee^{\ii\vc{k}\vdot\vc{r}}\dd[3]k\\&=-\frac{\ii}{\eta}[\vc{E}_\text{R}(\vc{r})-\vc{E}_\text{L}(\vc{r})]=\vc{H}_\text{R}(\vc{r})+\vc{H}_\text{L}(\vc{r})\,.\!\!\!\!
    \end{split}
\end{equation}
We can now see that fields $\vc{E}_{\text{R/L}}(\vc{r})$ and $\vc{H}_{\text{R/L}}(\vc{r})$ are superpositions of purely right-/left-handed modes which satify $\vc{H}_\text{R/L} = \mp \ii \vc{E}_\text{R/L}/\eta$. As we have already discussed in \cite{golat2024electromagnetic}, we can ---instead of keeping track of the electric field and the magnetic field--- choose the right-/left-handed basis defined by:
\begin{equation}\label{eq:FRL}
    \vc{F}_\text{R/L}(\vc{r})=\frac{\sqrt{\varepsilon}\vc{E}(\vc{r})\pm\ii\sqrt{\mu}\vc{H}(\vc{r})}{\sqrt{2}}=\frac{\sqrt{\varepsilon}\vc{E}_\text{R/L}(\vc{r})\pm\ii\sqrt{\mu}\vc{H}_\text{R/L}(\vc{r})}{\sqrt{2}}
\end{equation}
where the second equality comes from simply plugging in \cref{eq:ERL,eq:HRL}. One can also re-write Maxwell's equations in terms of these fields to find that the dynamics of the two fields completely decouple:
\begin{equation}
    \begin{split}
        \div\vc{F}_\text{R/L}=0\,,\quad
        \curl\vc{F}_\text{R/L}=\pm k\vc{F}_\text{R/L}\,.
    \end{split}
\end{equation}
Perhaps more surprisingly, several quadratic quantities (which would not usually be linear with fields) can also be seen as the linear sum of the corresponding quadratic quantities for each projection, such as the energy density:
\begin{equation}
    W = \frac{1}{4}({\varepsilon}\abs{\vc{E}}^2 + {\mu}\abs{\vc{H}}^2)= \frac{1}{4}(\abs{\vc{F}_\text{R}}^2 + \abs{\vc{F}_\text{L}}^2)=W_\text{R}+W_\text{L}\,,
\end{equation}%
the helicity density
\begin{equation}
    \mathfrak{S} = -\frac{1}{2\omega}\Im( \sqrt{\varepsilon\mu}{\vc{E}^*\!\vdot\vc{H}} )= \frac{1}{4\omega}(\abs{\vc{F}_\text{R}}^2 - \abs{\vc{F}_\text{L}}^2)=\mathfrak{S}_\text{R}+\mathfrak{S}_\text{L}=\frac{1}{\omega}(W_\text{R}-W_\text{L})\,,
\end{equation}%
the real poynting vector
\begin{equation}\label{eq:poyntsi}
    \Re\vc{\Pi}=\frac{1}{2}\Re(\vc{E}\cp\vc{H}^* )=\frac{c}{4}\Im(\vc{F}_\text{R}^*\cp\vc{F}_\text{R} - \vc{F}_\text{L}^*\cp\vc{F}_\text{L})=\Re\vc{\Pi}_\text{R}+\Re\vc{\Pi}_\text{L}\,,
\end{equation}%
and the spin angular momentum density
\begin{equation}\label{eq:spinsi}
    \vc{S}=\frac{1}{4\omega}\Im(\varepsilon{\vc{E}^*\!\cp\vc{E}} + \mu{\vc{H}^*\!\cp\vc{H}})=\frac{1}{4\omega}\Im(\vc{F}_\text{R}^*\cp\vc{F}_\text{R} + \vc{F}_\text{L}^*\cp\vc{F}_\text{L})=\vc{S}_\text{R}+\vc{S}_\text{L}=\frac{1}{\omega c}(\Re\vc{\Pi}_\text{R}-\Re\vc{\Pi}_\text{L})\,.
\end{equation}%
All of these can be easily confirmed using \cref{eq:FRL} to see that all the cross-terms vanish. The big advantage of this right-left basis over the usual electric-magnetic basis and more unusual parallel-antiparallel (introduced in \cite{golat2024electromagnetic}) is that the real Poynting vector can be decomposed in right and left components (\cref{eq:poyntsi}). In contrast, the Poynting vector for pure electric, magnetic, parallel, or antiparallel fields would be always zero, and only their linear combination will lead to non-zero energy flux. Thanks to this property of the right-left basis, one can consider the energy flux of each projection separately, and the two contributions are easy to compute (adding or subtracting \cref{eq:poyntsi,eq:spinsi}):
\begin{equation}\label{eq:fluxRL}
    \Re\vc{\Pi}_\text{R/L}=\frac{1}{2}(\Re\vc{\Pi}\pm\omega c \vc{S})\,.
\end{equation}
We can also appreciate the fact that for these fields we always have the poynting vector (kinetic momentum) and the spin either aligned or anti-aligned:
\begin{equation}
    \Re\vc{\Pi}_\text{R/L}=\pm\omega c \vc{S}_\text{R/L}\,,
\end{equation}
we know this to be true for right-/left-handed plane waves but it is true in general. The key difference is that for plane waves, the canonical momentum $\vc{p}$ and the kinetic momentum $\Re\vc{\Pi}/c^2$ coincide; however, this is not true in general. The canonical and kinetic momenta are related by the Belinfante-Rosenfeld relation
% \begin{equation}
%     \vc{p}_0=k\vc{S}_3-\frac{1}{2}\curl \vc{S}_0
% \end{equation}
% \begin{equation}
%     \vc{p}_3=k\vc{S}_0-\frac{1}{2}\curl \vc{S}_3
% \end{equation}
% \begin{equation}
%     (\vc{p}_\text{R}\pm\vc{p}_\text{L})=\frac{1}{c^2}(\Re\vc{\Pi}_\text{R}\pm\Re\vc{\Pi}_\text{L})-\frac{1}{2}\curl(\vc{S}_\text{R}\pm\vc{S}_\text{L})
% \end{equation}
\begin{equation}
    \vc{p}_\text{R/L}=\frac{1}{c^2}\Re\vc{\Pi}_\text{R/L}-\frac{1}{2}\curl\vc{S}_\text{R/L}\,,
\end{equation}
and this relation holds for each handedness individually.

Since in \cref{app:operatorPHFG} we obtained basis-independent versions of all the quantities studied in the main text, we can ask ourselves about their chiral basis expressions, as we will find they are particularly revealing. The dipoles themselves can also be written in the chiral basis, like the fields, as follows:
\begin{equation}
    \vc{\psi}(\vc{r})=\frac{A}{2\sqrt{\hbar\omega}}[\vc{F}_\text{R}(\vc{r}) \otimes\uv{e}_\text{R} +  \vc{F}_\text{L}(\vc{r})\otimes\uv{e}_\text{L}]\,,\quad
    \vc{\pi}(\vc{r})=\frac{A}{2\sqrt{\hbar\omega}}[\vc{\pi}_\text{R}(\vc{r}) \otimes\uv{e}_\text{R} +  \vc{\pi}_\text{L}(\vc{r})\otimes\uv{e}_\text{L}]\,,
\end{equation}
where the right-/left-handed dipoles are defined as: 
\begin{equation}
    \vc{\pi}_\text{R/L}(\vc{r})=\frac{\vc{p}(\vc{r})/\sqrt{\varepsilon}\pm\ii\sqrt{\mu}\vc{m}(\vc{r})}{\sqrt{2}}
\end{equation}
which plugged back into the basis independent expressions leads to power, helicity rate, force, and torque:
\begin{alignat}{2}
    \mathcal{P}_\text{abs}&=\frac{\omega}{2}\Im(\vc{F}_\text{R}^*\vc{\pi}_\text{R}+\vc{F}_\text{L}^*\vc{\pi}_\text{L})-\frac{\omega k^3}{12\pi}(\abs{\vc{\pi}_\text{R}}^2+\abs{\vc{\pi}_\text{L}}^2)\,,
\\
\omega\dot{\mathcal{H}}_\text{abs}&=\frac{\omega}{2}\Im(\vc{F}_\text{R}^*\vc{\pi}_\text{R}-\vc{F}_\text{L}^*\vc{\pi}_\text{L})-\frac{\omega k^3}{12\pi}(\abs{\vc{\pi}_\text{R}}^2-\abs{\vc{\pi}_\text{L}}^2)\,,
\\
{\vc{F}}&=\frac{1}{2}\Re\big[\vc{\pi}_\text{R}^\ast\vdot(\grad)\vc{F}_\text{R}+\vc{\pi}_\text{L}^\ast\vdot(\grad)\vc{F}_\text{L}\big]-\frac{k^4}{24\pi}\Im(\vc{\pi}_\text{R}^\ast\cp\vc{\pi}_\text{R}-\vc{\pi}_\text{L}^\ast\cp\vc{\pi}_\text{L}),
\\
{\vc{\varGamma}}&=\frac{1}{2}\Re(\vc{F}_\text{R}^*\cp\vc{\pi}_\text{R}+\vc{F}_\text{L}^*\cp\vc{\pi}_\text{L})
    -\frac{ k^3}{12\pi}\Im\big(\vc{\pi}_\text{R}^*\cp\vc{\pi}_\text{R}+\vc{\pi}_\text{L}^*\cp\vc{\pi}_\text{L}\big).
\end{alignat}
We can see that all of these quantities acquire a rather simple form in this basis. This basis is not only simpler for these quantities but also insightful; for example, consider the quantity
\begin{alignat}{2}
    \frac{1}{2}(\mathcal{P}_\text{abs}\pm\omega\dot{\mathcal{H}}_\text{abs})&=\frac{\omega}{2}\Im(\vc{F}_\text{R/L}^*\vc{\pi}_\text{R/L})-\frac{\omega k^3}{12\pi}\abs{\vc{\pi}_\text{R/L}}^2\,,
\end{alignat}
which seems to only depend on fields and dipoles of single handedness. The absorbed power is calculated from the total energy flux (that is considering both incident and scattered ones):
\begin{equation}
    \mathcal{P}_\text{abs}=\iiint\Re\vc{\Pi}_\text{tot}\vdot\dd{s}\,,
\end{equation}
the helicity density is defined in a similar way as an integral of the helicity flux which is $c\vc{S}_\text{tot}$ 
\begin{equation}
    \dot{\mathcal{H}}_\text{abs}=\iiint c\vc{S}_\text{tot}\vdot\dd{s}\,,
\end{equation}
therefore, it is clear from \cref{eq:fluxRL} that the quantity represents the absorbed power of each handedness:
\begin{equation}
    \mathfrak{P}_\text{abs}^\text{R/L}=\frac{1}{2}(\mathcal{P}_\text{abs}\pm\omega\dot{\mathcal{H}}_\text{abs})=\frac{1}{2}\iiint(\Re\vc{\Pi}_\text{tot}\pm\omega c \vc{S}_\text{tot})\vdot\dd{s}=\iiint\Re\vc{\Pi}^\text{tot}_\text{R/L}\vdot\dd{s}\,,
\end{equation}
because $\Re\vc{\Pi}^\text{tot}_\text{R/L}$ is the total energy flux for each handedness (again considering both incident and scattered fields). In this sense, the helicity absorption rate is proportional to the difference between the right- and left-handed absorptions:
\begin{equation}
    \dot{\mathcal{H}}_\text{abs}=\frac{1}{\omega}\iiint\Re(\vc{\Pi}^\text{tot}_\text{R}-\vc{\Pi}^\text{tot}_\text{L})\vdot\dd{s}\,,
\end{equation}
whilst the total absorbed power is their sum 
\begin{equation}
    {\mathcal{P}}_\text{abs}=\iiint\Re(\vc{\Pi}^\text{tot}_\text{R}+\vc{\Pi}^\text{tot}_\text{L})\vdot\dd{s}\,.
\end{equation}
Quantities $\mathfrak{P}_\text{abs}^\text{R/L}$, $\mathfrak{P}_\text{ext}^\text{R/L}$, $\mathfrak{P}_\text{sca}^\text{R/L}$ are in principle measurable by filtering either handedness at the output, and it would represent handedness resolved power.

% \begin{equation*}
%    \frac{\omega}{2}\big[{\Im\qty(\vc{E}^\ast\vdot\vc{p}+\mu\vc{H}^\ast\vdot\vc{m})}-{\frac{k^3}{6\pi}(\frac{1}{\varepsilon}\abs{\vc{p}}^2+\mu\abs{\vc{m}}^2)}\big]
% \end{equation*}
% \begin{equation*}
%    \frac{\omega}{2}\big[{\Im\qty(\vc{E}^\ast\vdot\vc{p}-\mu\vc{H}^\ast\vdot\vc{m})}-{\frac{k^3}{6\pi}(\frac{1}{\varepsilon}\abs{\vc{p}}^2-\mu\abs{\vc{m}}^2)}\big]
% \end{equation*}
% \begin{equation*}
%    \frac{\omega}{2}\big[\,\Im(\vc{F}_\text{R}^*\vc{\pi}_\text{R}-\vc{F}_\text{L}^*\vc{\pi}_\text{L})-\frac{k^3}{6\pi}(\abs{\vc{\pi}_\text{R}}^2-\abs{\vc{\pi}_\text{L}}^2)\big]
% \end{equation*}
% \begin{equation*}
%    \frac{\omega}{2}\big[\,\Im(\vc{F}_\text{a}^*\vc{\pi}_\text{a}-\vc{F}_\text{p}^*\vc{\pi}_\text{p})-\frac{k^3}{6\pi}(\abs{\vc{\pi}_\text{a}}^2-\abs{\vc{\pi}_\text{p}}^2)\big]
% \end{equation*}
\subsection{Cross sections for handedness resolved power in the right-left basis}
The handedness resolved power $\mathfrak{P}_\text{R/L}$ (that is, the power absorbed, extinguish or scattered of a given helicity measured at the detector, for any incident illuminating polarisation, not to be confused with the power absorbed, extinguished or scattered of any handedness under a given single-handed circularly polarised illumination $\mathcal{P}^\text{R/L}$) can also be written in terms of energy densities and cross sections:
\begin{equation}
    \mathfrak{P}_\text{R/L}=\tfrac{1}{2}(\mathcal{P}\pm\omega\dot{\mathcal{H}})=\sum_{A=0}^3\tfrac{1}{2}(\sigma_A\pm\gamma_A)cW_A\,.
\end{equation}
Combining the power and helicity extinction cross sections, we see that they can be expressed in the right-left basis
\begin{alignat}{3}
    \tfrac{1}{2}(\sigma_\text{ext}^0\pm\gamma_\text{ext}^0)&=\tfrac{k}{2}\Im({\alpha_0\pm\alpha_3})&&=k\Im{\alpha_\text{R/L}}\,,\\
    \tfrac{1}{2}(\sigma_\text{ext}^1\pm\gamma_\text{ext}^1)&=\tfrac{k}{2}\Im({\alpha_1\mp\ii\alpha_2})&&=k\Im{\alpha_\text{RL/LR}}\,,\\
    \tfrac{1}{2}(\sigma_\text{ext}^2\pm\gamma_\text{ext}^2)&=\tfrac{k}{2}\Im({\alpha_2\pm\ii\alpha_1})&&=\pm k\Re{\alpha_\text{RL/LR}}\,,\\
    \tfrac{1}{2}(\sigma_\text{ext}^3\pm\gamma_\text{ext}^3)&=\tfrac{k}{2}\Im({\alpha_3\pm\alpha_0})&&=\pm k\Im{\alpha_\text{R/L}}
\end{alignat}
where one can therefore identify 
$\alpha_\text{RL/LR}=(\alpha_1\mp\ii\alpha_2)/2$, 
$\alpha_\text{LR}+\alpha_\text{RL}=\alpha_1$, 
$\ii\alpha_\text{RL}-\ii\alpha_\text{LR}=\alpha_2$ and 
$\alpha_{R/L}=(\alpha_0\pm\alpha_3)/2$. 
These four elements are the components of a $ 2 \times 2 $ polarisability matrix in the right-left basis:
\begin{equation}
\begin{alignedat}{3}
\begin{pmatrix}
\vc{\pi}_\text{R}\\
\vc{\pi}_\text{L}
\end{pmatrix}=
\begin{pmatrix}
\alpha_\text{R} & \alpha_\text{RL} \\
\alpha_\text{LR} & \alpha_\text{L}
\end{pmatrix} 
\begin{pmatrix}
\vc{F}_\text{R}\\
\vc{F}_\text{L}
\end{pmatrix}
= \begin{pmatrix}
\alpha_0 + \alpha_3 & \alpha_1 - \ii \alpha_2 \\
\alpha_1 + \ii \alpha_2 & \alpha_0 - \alpha_3
\end{pmatrix}\,.
\end{alignedat}
\end{equation}%
Similarly, the scattering cross sections are:
\begin{alignat}{3}
    \tfrac{1}{2}(\sigma_\text{sca}^0\pm\gamma_\text{sca}^0)&=
    \tfrac{k^4}{6\pi}(\abs*{\alpha_\text{R/L}}^2+\abs{\alpha_\text{RL/LR}}^2)
    \,,\\
    \tfrac{1}{2}(\sigma_\text{sca}^1\pm\gamma_\text{sca}^1)&=
    \tfrac{k^4}{3\pi}\Re(\alpha_\text{R/L}^*\alpha_\text{RL/LR})
    \,,\\
    \tfrac{1}{2}(\sigma_\text{sca}^2\pm\gamma_\text{sca}^2)&=
    \pm \tfrac{k^4}{3\pi}\Im(\alpha_\text{R/L}^*\alpha_\text{LR/RL})
    \,,\\
    \tfrac{1}{2}(\sigma_\text{sca}^3\pm\gamma_\text{sca}^3)&=
    \pm \tfrac{k^4}{6\pi}(\abs*{\alpha_\text{R/L}}^2-\abs{\alpha_\text{RL/LR}}^2)\,.
\end{alignat}

We can again look at the paraxial limit with \( W_0 = W_\text{R} + W_\text{L} \), \( W_3 = W_\text{R} - W_\text{L} \) and \( W_1 = W_2 = 0 \):
\begin{equation}
\begin{split}
    \begin{pmatrix}
        \mathfrak{P}^{\mathrm R}_{\mathrm{ext}} \\
        \mathfrak{P}^{\mathrm L}_{\mathrm{ext}}
    \end{pmatrix}
    &=
    2k
    \begin{pmatrix}
        \Im{\alpha_{\mathrm R}} & 0 \\
        0 & \Im{\alpha_{\mathrm L}}
    \end{pmatrix}
    \begin{pmatrix}
        cW_{\mathrm R} \\
        cW_{\mathrm L}
    \end{pmatrix},\qq{where}
     \mathfrak{P}^\text{R/L}_\text{ext}=\tfrac{1}{2}(\mathcal{P}_\text{ext}\pm\omega\dot{\mathcal{H}}_\text{ext})\,.
\end{split}
\end{equation}
% \begin{equation*}
% % \begin{split}
%     \mathfrak{P}^\text{R/L}_\text{ext}=\tfrac{1}{2}(\mathcal{P}_\text{ext}\pm\omega\dot{\mathcal{H}}_\text{ext})=\sum_{A=0}^3\tfrac{1}{2}(\sigma_\text{ext}^A\pm\gamma_\text{ext}^A)cW_A=(2k\Im\alpha_\text{R/L}
%    )cW_\text{R/L}=\sigma_\text{ext}^\text{R/L}
%    cW_\text{R/L}=\mathcal{P}^\text{R/L}_\text{ext}
%     \,,
% % \end{split}
% \end{equation*}
from which we can see that in the paraxial limit, the right- or left-handed helicity resolved extinguished power is the same as if we measured the power with only the right- or left-handed illumination. In other words, for extinction, it does not matter if we filter one of the two circularly polarised polarisations at the input or at the output. However, this is not true for absorption or scattering. For scattering we have:
% \begin{equation}\label{eq:handedpower}
% \begin{split}
%     \mathfrak{P}^\text{R}_\text{sca}=\tfrac{1}{2}(\mathcal{P}_\text{sca}+\omega\dot{\mathcal{H}}_\text{sca})&=\sum_{A=0}^3\tfrac{1}{2}(\sigma_\text{sca}^A+\gamma_\text{sca}^A)cW_A=
%     \tfrac{k^4}{3\pi}c\big(\abs*{\alpha_\text{R}}^2W_\text{R}+\abs*{\alpha_\text{RL}}^2W_\text{L}\big)
%     \,,\\
%     \mathfrak{P}^\text{L}_\text{sca}=\tfrac{1}{2}(\mathcal{P}_\text{sca}-\omega\dot{\mathcal{H}}_\text{sca})&=\sum_{A=0}^3\tfrac{1}{2}(\sigma_\text{sca}^A-\gamma_\text{sca}^A)cW_A=
%     \tfrac{k^4}{3\pi}c\big(\abs*{\alpha_\text{L}}^2W_\text{L}+\abs*{\alpha_\text{LR}}^2W_\text{R}\big)
%     \,.
% \end{split}
% \end{equation}
\begin{equation}\label{eq:handedpower}
\begin{split}
    \begin{pmatrix}
        \mathfrak{P}^{\mathrm R}_{\mathrm{sca}} \\
        \mathfrak{P}^{\mathrm L}_{\mathrm{sca}}
    \end{pmatrix}
    &=
    \frac{k^4}{3\pi}
    \begin{pmatrix}
        \abs{\alpha_{\mathrm R}}^2 & \abs{\alpha_{\mathrm{RL}}}^2 \\
        \abs{\alpha_{\mathrm{LR}}}^2 & \abs{\alpha_{\mathrm L}}^2
    \end{pmatrix}
    \begin{pmatrix}
        cW_{\mathrm R} \\
        cW_{\mathrm L}
    \end{pmatrix},\qq{where}
     \mathfrak{P}^\text{R/L}_\text{sca}=\tfrac{1}{2}(\mathcal{P}_\text{sca}\pm\omega\dot{\mathcal{H}}_\text{sca})\,.
\end{split}
\end{equation}
% If we take right/left-handed illumination with $W_0=\pm W_3=W_\text{R/L}$, then the scattered power will be
% \begin{equation*}
%     \mathcal{P}^\text{R/L}_\text{sca}=\tfrac{k^4}{3\pi}(\abs*{\alpha_\text{R/L}}^2+\abs{\alpha_\text{LR/RL}}^2)cW_\text{R/L}=\mathfrak{P}^\text{R/L}_\text{sca}+\mathfrak{P}^\text{L/R}_\text{sca}\,,
% \end{equation*}
% but from \cref{eq:handedpower} we know that only the first term has the same handedness as the illumination, because for this illumination the helicity resolved powers will be
% \begin{equation*}
% \begin{split}
%     \mathfrak{P}^\text{R/L}_\text{sca}=\tfrac{1}{2}(\mathcal{P}^\text{R/L}_\text{sca}\pm\omega\dot{\mathcal{H}}^\text{R/L}_\text{sca})&=
%     \tfrac{k^4}{3\pi}\abs*{\alpha_\text{R/L}}^2cW_\text{R/L}
%     \,,\\
%     \mathfrak{P}^\text{L/R}_\text{sca}=\tfrac{1}{2}(\mathcal{P}^\text{R/L}_\text{sca}\mp\omega\dot{\mathcal{H}}^\text{R/L}_\text{sca})&=
%     \tfrac{k^4}{3\pi}\abs*{\alpha_\text{LR/RL}}^2cW_\text{R/L}
%     \,.
% \end{split}
% \end{equation*}
What this means physically is that particles with $\abs*{\alpha_\text{LR/RL}}\neq0$ change the helicity of the incident light when they scatter it, so while the input polarisation has only one helicity, the scattered light will have both. And since 
\begin{equation}
    \begin{pmatrix}
        \mathfrak{P}^{\mathrm R}_{\mathrm{abs}} \\
        \mathfrak{P}^{\mathrm L}_{\mathrm{abs}}
    \end{pmatrix}
    =
    \begin{pmatrix}
        \mathfrak{P}^{\mathrm R}_{\mathrm{ext}} \\
        \mathfrak{P}^{\mathrm L}_{\mathrm{ext}}
    \end{pmatrix}
    -
    \begin{pmatrix}
        \mathfrak{P}^{\mathrm R}_{\mathrm{sca}} \\
        \mathfrak{P}^{\mathrm L}_{\mathrm{sca}}
    \end{pmatrix}
\end{equation}
the same will be true for absorbed power.

\section{Measuring cross sections}\label{app:measuring}
If we want to find $\sigma_0$ and $\sigma_3$ (for extinction/scattering or absorption) all we have to do is illuminate the dipole separately by the right- and left-handed paraxial radiation with the same intensity ${cW_\text{R}=cW_\text{L}=cW}$ and measure the extinct/scattered or absorbed power, one can then use \cref{eq:power_paraxial} to show that 
\begin{align}
    \sigma_0=\frac{1}{2}\frac{\mathcal{P}_\text{R}+\mathcal{P}_\text{L}}{cW}=\frac{1}{2}\qty(\sigma_\text{R}+\sigma_\text{L})\,,\quad
    \sigma_3=\frac{1}{2}\frac{\mathcal{P}_\text{R}-\mathcal{P}_\text{L}}{cW}=\frac{1}{2}\qty(\sigma_\text{R}-\sigma_\text{L})\,,
\end{align}
where $\mathcal{P}_\text{R/L}$ are the measured powers under right- and left-handed illumination.
In the case of extinction, the total cross section $\sigma_\text{ext}^0$ and its chiral counterpart $\sigma_\text{ext}^3$ are usually related by the so-called dissymmetry factor or $g$-factor, which is defined as the ratio:
\begin{equation}
    g\equiv\frac{\mathcal{P}_\text{ext}^\text{R}-\mathcal{P}_\text{ext}^\text{L}}{(\mathcal{P}_\text{ext}^\text{R}+\mathcal{P}_\text{ext}^\text{L})/2}=2\frac{\sigma_\text{ext}^3}{\sigma_\text{ext}^0}\,.
\end{equation}
If the particles are illuminated by near-field radiation (such as evanescent waves), the quantities $W_1$ and $W_2$ no longer need to be zero. For example, in the case of evanescent waves, one can control which of the $W_1$, $W_2$ and $W_3$ are non-zero. Consider an evanescent wave with effective index $n=k_x/k$ (associated to $\kappa=\sqrt{n^2-1}$) such that $\vc{k}=k(n \hat{\vc{x}} + \ii \kappa \hat{\vc{z}})$, with arbitrary polarisation:
\begin{equation}\label{EHevanescent}
    \sqrt{\varepsilon}\vc{E}=\pqty{
        \ii \kappa A_p,A_s,-n A_p 
    }^\intercal \ee^{\ii \vc{k} \cdot \vc{r} }, \quad
    \sqrt{\mu}\vc{H}=\pqty{
        -\ii \kappa A_s,A_p,n A_s 
    }^\intercal\ee^{\ii \vc{k} \cdot \vc{r} },
\end{equation}
where $e^{\ii \vc{k} \cdot \vc{r} } = e^{\ii k n x - k \kappa z}$ such that a plane wave is recovered when $(n=1, \kappa=0)$. The amplitudes $A_p$ and $A_s$ correspond to the usual p-polarised and  s-polarised components, which can be seen as vertical and horizontal polarisation. 
The `energy densities' will be proportional to the Stokes parameters in the following way \cite{golat2024electromagnetic}:
\begin{alignat*}{4}
W_0 &=\tfrac{1}{2} e^{-2 \kappa z} n^2 \mathcal{S}_0,%\\
&\quad
W_2  &=\tfrac{1}{2}e^{-2 \kappa z} \kappa^2 \mathcal{S}_2\,, 
% \\
&\quad
W_1&=\tfrac{1}{2} e^{-2 \kappa z} \kappa^2 \mathcal{S}_1\,,%\\
&\quad
W_3  &=\tfrac{1}{2} e^{-2 \kappa z} n^2 \mathcal{S}_3\,,
\end{alignat*}
where the Stokes parameters can be defined using the amplitudes: 
\begin{alignat*}{4}
\mathcal{S}_0 &=\abs{A_p}^2+\abs{A_s}^2,%\\
&\quad
\mathcal{S}_2  &=2 \Re(A_p^*A_s)=\abs{A_\text{d}}^2-\abs{A_\text{a}}^2\,, 
% \\
&\quad
\mathcal{S}_1&=\abs{A_p}^2-\abs{A_s}^2\,,%\\
&\quad
\mathcal{S}_3  &=2 \Im(A_p^*A_s)=\abs{A_\text{R}}^2-\abs{A_\text{L}}^2\,.
\end{alignat*}
Using \cref{eq:dipolar_power} it is possible to show that the remaining cross sections can be obtained from the differences of powers measured for different polarisations (either horizontal-vertical or diagonal-antidiagonal) of the evanescent wave:
\begin{align}
    \sigma_1=\frac{\kappa^2}{2n^2}\frac{\mathcal{P}_p-\mathcal{P}_s}{cW}=\frac{1}{2}\qty(\sigma_\text{e}-\sigma_\text{m})\,,
    % \\
    \quad
    \sigma_2=\frac{\kappa^2}{2n^2}\frac{\mathcal{P}_\text{d}-\mathcal{P}_\text{a}}{cW}=\frac{1}{2}\qty(\sigma_\text{a}-\sigma_\text{p})\,.
\end{align}
Note that factor $\kappa^2/n^2=(\sin^2\theta_\text{i}-\sin^2\theta_\text{c})/\sin^2\theta_\text{i}$, where $\theta_\text{i}$ is the angle of incidence and $\theta_\text{c}$ is the critical angle. The same should be in principle possible for helicity using \cref{eq:dipolar_helicity}.

\end{document}